# Multi-rater delta: extending the delta nominal measure of agreement between two raters to many raters

*Martín Andrés, A.*[1(1)] *and Álvarez Hernández, M.*[(2)]

(1) Bioestadística, Facultad de Medicina, Universidad de Granada. Granada, Spain.

(2) Centro Universitario de la Defensa – Escuela Naval Militar, Marín (Pontevedra), Spain.

## ABSTRACT

The need to measure the degree of agreement among *R* raters who independently classify *n* subjects within *K* nominal categories is frequent in many scientific areas. The most popular measures are Cohen's *kappa* ($R$=2), Fleiss' *kappa*, Conger's *kappa* and Hubert's *kappa* ($R$≥2) coefficients, which have several limitations. In 2004, the *delta* coefficient was defined for the case of $R=2$, which did not have the limitations of Cohen's *kappa* coefficient. This article extends the coefficient *delta* from $R$=2 raters to $R$≥2 (coefficient *multi-rater delta*), demonstrating that it can be expressed in the kappa format and has the same advantages as the coefficient *delta* with regard to the type *kappa* classic coefficients: *i)* it refers to the proportion of replies that are concordant not by chance; *ii)* allow to obtain a parameter that faithfully measures the degree of agreement in each category; and *iii)* it is not affected by the marginal imbalance.

**Keywords:** Cohen's *kappa*; Conger's *kappa*; *Delta* agreement; Fleiss' *kappa*; Hubert's *kappa*; nominal agreement.

## 1. Introduction

In many fields of science, including the behavioural sciences, geography and medicine, the degree of concordance or agreement among *R* raters that independently classify *n* subjects within *K* unordered categories is assessed (Fleiss 1971, Landis & Koch 1975a and b, Warrens 2010, Schuster & Smith 2005). In this article, we focus on the case in which no gold-standard rater exists.

[1] Correspondence to: Bioestadística. Facultad de Medicina, C8-01. Universidad de Granada. 18071 Granada. Spain. E-mail: amartina@ugr.es. Phone:34-58-244080. Fax: 34-58-246117.

Let us consider the case of two raters ($R=2$). Because some of the observed agreements may occur due to chance, the most common action is to eliminate the effect of chance using Cohen's *kappa* coefficient ($\kappa_C$) (Cohen 1960). Although *Kappa* is a very popular and easily calculated measure of agreement, it has several limitations (Brennan & Prediger 1981, Agresti *et al.* 1995, Guggenmoos-Holzmann & Vonk 1998, Nelson & Pepe 2000, Martín Andrés & Femia Marzo 2005, Erdmann *et al.* 2015). The two most relevant limitations are its dependence on marginal distributions and the difficulty in measuring the degree of agreement for each category. Martín Andrés & Femia Marzo (2004, 2005 and 2008) proposed a response model (*delta* model) that led to the measure of agreement *delta* ($\Delta$). Because it does not have the limitations of $\kappa_C$ (Ato *et al.* 2011; Shankar and Bangdiwala 2014, Giammarino *et al.* 2021), it has been occasionally used in many different fields (Ecology, Geography, Psychology, Medicine, …). The *delta* model employs several measures that are valid in all circumstances, even if no gold-standard exists or if there exists, its marginal distribution is not fixed. However, the *delta* model is based on the assumption that one of the two raters is a gold-standard, which explains why its parameters are directly related to the situation. The first aim of this article is redefine the model for the case in which neither rater is a gold-standard. In addition, this extension will identify some minor errors committed by Martín Andrés & Femia Marzo when estimating the parameters of their *delta* model. A traditional example is Table 1(a), which comes from the classic example of Fleiss *et al.* (2003) where two raters diagnose 100 subjects in $K$=3 categories (Psychotic, Neurotic and Organic).

Now, let us consider the case where there are many raters ($R\geq 2$). Various authors have proposed generalizations of $\kappa_C$ for the multi-rater case (Fleiss 1971, Hubert 1977, Conger 1980, Warrens 2010, Marasini *et al.* 2016). Although Fleiss' *kappa* ($\kappa_F$) is the most popular measure, it is not an extension of $\kappa_C$ since the value of $\kappa_F$ is not equal to $\kappa_C$ when $R$=2 (Conger 1980); however, it is an extension of Scott's *pi* ($\pi$) coefficient (Scott 1955) (see the quotations from Warrens 2010 and Marasini *et al.* 2016). Conversely, the generalizations of Hubert's *kappa* (Hubert, 1977) and Conger's *kappa* (Conger 1980) are an extension of $\kappa_C$, particularly the two generalizations which will be noted later as $\kappa_{H2}$ (Hubert *pairwise*) and $\kappa_{HR}$ (Hubert *R-wise*). All *multi-rater kappa* coefficients exhibit a paradoxical behaviour (Marasini *et al.* 2016, Conger 2017), such as their dependence on the marginal imbalance. The second objective of this article is to extend the *delta* model from $R$=2 to $R$≥2 (*multi-rater delta* model). An example for $R$=$K$=3 is listed in Table 2(a), from Dillon & Mulani (1984); this example is cited by Schuster & Smith (2005) and "analyzed a persuasive communication study for which three raters classified 164 subjects as either positive, neutral, or negative"; however, in this paper, the ordinal quality will be analysed as if it were a nominal quality.

Additionally, in the *delta* model for $R$=2, the parameter estimates and their standard errors (SEs) are obtained by the score method, which generates complicated deductions and expressions. Obtaining simpler equivalent expressions by a less cumbersome method, such as the classical multivariate *delta* method, is the third aim of this article. These expressions will be extended to the case $R$≥2.

Finally, the current delta model should be framed within the scope of the known models. The analysis of the existing nominal agreement between several raters can be approached from various perspectives (Agresti 1992, Uebersax 1992, Banerjee *et al.* 1999 and Barlow 2005). The most convenient option is usually the latent class model, which considers the existence of unobserved or latent variables (Dillon and Mulani 1984, Uebersax and Grove1990, Klauer and Batchelder 1996). The usual latent class model assumes that the rating level assigned by one rater is statistically independent of the rating levels assigned by other raters. The current *delta* model is loosely related to the latent class model because its origin is in the model by Martín Andrés and Femia Marzo (2004), which in turn is derived from the model by Martín Andrés and Luna del Castillo (1989) for multiple choice tests (an extension of the classic model by Lord and Novick 1968). And precisely the models of Martín Andrés and Luna del Castillo (1989) and the latent classes of Klauer & Batchelder (1996) are formally the same.

This article is organized as follows: in Section 2, the classic models *kappa* and *delta* are introduced, as well as the new *delta* model for the case of $R$=2. In Section 3, three classic *kappa* models are introduced as well as the new *multi-rate delta* model, for the $R\geq 2$ case. In Section 4 the parameters of the *multi-rater delta* model and its standard errors (SE) are estimated. Section 5 offers several examples and finally, in Section 6, the conclusions are set out.

## 2. Models for two raters

Let us start with the case of two raters ($R=2$) that independently classify $n$ subjects within $K$ nominal categories. Given a subject, rater 1 classifies it as type $i$ ($i$ = 1, 2, ..., $K$) and rater 2 as type $j$ ($j$ = 1, 2, ..., $K$), which generates a table of absolute frequencies $x_{ij}$ and a table of relative frequencies (observed cell proportions) $\overline{p}_{ij}=x_{ij}/n$, with $\sum_{i=1}^{K}\sum_{j=1}^{K}x_{ij}=n$ and $\sum_{i=1}^{K}\sum_{j=1}^{K}\overline{p}_{ij}=1$. The notation for the observed totals of row ($x_{i\bullet}$ and $\overline{p}_{i\bullet}$), column ($x_{\bullet j}$ and $\overline{p}_{\bullet j}$) or overall ($x_{\bullet\bullet}=n$ and $\overline{p}_{\bullet\bullet}=1$) is typical; please refer to Table 1(a). If $p_{ij}$ is the probability that a subject is classified in cell ($i$, $j$), then the observed data set $\{x_{ij}\}$ is derived from a multinomial distribution of parameters $n$ and $\{p_{ij}\}$; $\{p_{i\bullet}\}$ and $\{p_{\bullet j}\}$ will be the marginal distributions of row raters and column raters, respectively.

To analyse the previous problem, Cohen (1960) defined the classic *kappa* coefficient $\kappa_C=(I_o-I_e)/(1-I_e)$ estimated by $\hat{\kappa}_C=\left(\overline{I}_o-\overline{I}_e\right)/\left(1-\overline{I}_e\right)$, where $I_o=\sum_{i=1}^{K}p_{ii}$ ($\overline{I}_o=\sum_{i=1}^{K}\overline{p}_{ii}$) is the real (estimated) observed agreement index, and $I_e = \sum_{i=1}^{K} p_{i\bullet} p_{\bullet i}$ ($\overline{I}_e = \sum_{i=1}^{K} \overline{p}_{i\bullet} \overline{p}_{\bullet i}$) is the real (estimated) expected agreement index; the estimated values have been obtained on the assumption of independence between the classifications of the two raters. The denominator $\left(1-\overline{I}_e\right)$ has the two following effects on $\hat{\kappa}_C$ (Martín Andrés & Femia Marzo 2004); similarly with $(1-I_e)$ for $\kappa_C$. On the one hand its value is very dependent on the marginal distributions. On the other, its interpretation is not absolute, but relative to the value of $\left(1-\overline{I}_e\right)$; for the example in Table 1(a) one obtains $\hat{\kappa}_C$=.6765, indicating that "the two raters agree on 67.65% of the maximum number possible of non-random agreements".

One way to correct these two limitations is by using the following *delta* model (Martín Andrés & Femia Marzo, 2004):

$$p_{ij}/p_{i\bullet} = \delta_{ij}\Delta_i + (1-\Delta_i)\pi_j\,, \qquad \textbf{(1)}$$

where $\delta_{ij}$ allude to the Kronecker delta, $0 \leq \pi_j \leq 1$, $\sum_{j=1}^{K} \pi_j = 1$ and $\Delta_i \leq 1$. This model is based on the model that Martín Andrés and Luna del Castillo (1989) used in the context of multiple choice tests −see their expression (3)− which in turn is based on the classic model by Lord and Novick (1968); the model of Martín Andrés and Luna del Castillo (1989) is the same as the model of the expression (1) of Klauer and Batchelder (1996), regardless of the name given to each parameter. This is why the response model by the current expression (1) can be interpreted as follows. When rater 2 faces a subject classified as type $i$ by rater 1 (who is assumed to be a gold-standard rater), s/he recognizes the subject as type $i$ with probability $\Delta_i$; when it is recognized, s/he correctly classifies it. When s/he does not recognize it, s/he does so with a probability $1-\Delta_i$, randomly classifies it as type $j$ with a probability $\pi_j$. Given that $p_{ii} = p_{i\bullet}\Delta_i + p_{i\bullet}(1-\Delta_i)\pi_i$, the proportion of agreements in category $i$ is the sum of the proportions of agreements that are not due to chance -the first summand $p_{i\bullet}\Delta_i$- and the proportions of agreements due to chance -the second summand $p_{i\bullet}(1-\Delta_i)\pi_i$. Hence, the total proportion of agreements not occurring by chance will be $\Delta=\Sigma p_{i\bullet}\Delta_i$. The authors (Martín Andrés & Femia Marzo 2005) found that the value of the estimator $\hat{\Delta}$ of $\Delta$ is usually very similar to the value of $\hat{\kappa}_C$, except in certain cases in which at least one marginal distribution is very unbalanced. In the example in Table 1(a) one obtains $\hat{\Delta}$=.6875, a very similar value to that of $\hat{\kappa}_C$=.6765 despite the imbalance of the two marginals, but in other cases the values can be very different (see the second paragraph of Section 5). However, the interpretation is now more direct: 68.75% of the responses are concordant beyond chance (as opposed to 89% of raw agreements).

It can be seen that the classic *delta* model is formulated under the idea that the row rater is a gold-standard, even when the model is also otherwise valid. To eliminate this dependence, in Appendix A it is proved that the classic *delta* model is equivalent to the following new *delta* model:

$$p_{ij} = \delta_{ij}\alpha_i + (1-\Delta)\pi_{i1}\pi_{j2} \tag{2}$$

where $\alpha_i \leq 1$, $\Delta=\Sigma\alpha_i\leq 1$ (which is the same parameter $\Delta$ as in the classic *delta* model), $0\leq\pi_{i1}\leq 1$, $0\leq\pi_{j2}\leq 1$ and $\sum_{i=1}^{K}\pi_{i1}=\sum_{j=1}^{K}\pi_{j2}=1$. The following response model is assumed in the new *delta* model (see Appendix A). When the two raters are faced with a given subject, both raters recognize it as category $i$ with a probability of $\alpha_i$; when they recognize it, they classify it as type $i$; when they do not recognize it, they do so with probability $1-\Delta$, classifying it randomly and independently with probability distributions $\{\pi_{i1}\}$ and $\{\pi_{j2}\}$, respectively.

If no model is considered, there exist two "raw" parameters of interest. The first is the total proportion of agreements $p=\sum_{i=1}^{K}p_{ii}$, where $p_{ii}$ is the proportion of agreements in category $i$. The parameter $p_{ii}$ does not measure the degree of agreement in category $i$, but does measure the contribution of category $i$ to the total agreement, since $p_{ii}$ is dependent on the degree of agreement in category $i$ and the marginal distributions $p_{i\bullet}$ and $p_{\bullet i}$ in category $i$; for example, if $p_{ii}$= $p_{i\bullet}$ =$p_{\bullet i}$=0.01, then the degree of agreement in class $i$ is maximal, but the proportion of agreements ($p_{ii}$=0.01) is very small due to the fact that the marginal proportions are also very small. A suitable method for defining the degree of agreement in category $i$ (the second parameter) is through the *consistency* $S_i=2p_{ii}/(p_{i\bullet}+p_{\bullet i})$ of Martín Andrés & Femia Marzo (2005), which is a parameter that is based on the "proportion of specific agreement" of Fleiss *et al.* (2003) and on the agreement index of Cichetti & Feinstein (1990). Note that $S_i$ is the proportion of agreements obtained among all of the responses $i$ of any one of the two raters. The two agreement parameters $p$ and $S_i$ are raw parameters, since they are defined without considering the effect of chance. Since $p_{ii}=\alpha_i+(1-\Delta)\pi_{i1}\pi_{i2}$ from expression (2), then $p_{ii}$ is the sum of the proportion of agreements $\alpha_i$ that do not occur by chance and the proportion of agreements $(1-\Delta)\pi_{i1}\pi_{i2}$ that do occur by chance. If in the two previous raw parameters $p_{ii}$ are replaced by $\alpha_i$, then two parameters that are corrected by chance will be obtained (which is the current objective). In the new *delta* model for two raters, the consequence is that the two parameters of interest are *the overall degree of agreement* $\Delta=\Sigma\alpha_i$ (that is, the total proportion of agreements that do not occur by chance), and the *consistency* $\mathcal{S}_i$ in category $i$ (that is, the degree of agreement in category $i$ that do not occur by chance),

$$\mathcal{S}_i=\frac{2\alpha_i}{p_{i\bullet}+p_{\bullet i}}=\frac{2\alpha_i}{2\alpha_i+(1-\Delta)(\pi_{i1}+\pi_{i2})}, \qquad \textbf{(3)}$$

where $\alpha_i$ is the proportion of agreement in category $i$ that do not occur by chance and the second equality is due to $p_{i\bullet}=\sum_{j=1}^{K}p_{ij}=\alpha_i+(1-\Delta)\pi_{i1}$ and $p_{\bullet j}=\sum_{i=1}^{K}p_{ij}=\alpha_j+(1-\Delta)\pi_{j2}$. In the new *delta* model, note that the possibility that $\mathcal{S}_i<0$, $\alpha_i<0$ or $\Delta<0$ is allowed because the agreement can sometimes be negative. This can be interpreted as one of the raters classifying subjects "the other way around" compared to the other rater; that is, if the subject is in category $i$ for one of the raters, the subject is in a category other than $i$ for the other rater.

Martín Andrés & Femia Marzo (2005) explained that $\mathcal{S}_i$ has the same objective as that pursued when defining $\kappa_C$ for collapsed data in category $i$ -parameter $\kappa_{C(i)}$- which is an aim that is always achieved with $\mathcal{S}_i$; however, sometimes it is not achieved with $\kappa_{C(i)}$. In fact, a total agreement parameter based on the collapsed data in category $i$ measures the degree of total agreement in the new situation, but does not measure the degree of agreement in category $i$ because it should also measure the degree of agreement in the "not $i$" category.

**3. Models for many raters**

Let there now be $R\geq 2$ raters who independently classify $n$ subjects in $K$ categories, producing a data matrix $\{y_{sr}\}$, with $n$ rows ($s$=1, 2, …, $n$), $R$ columns ($r$=1, 2, …, $R$) and values $y_{sr}$=1, 2, …, $K$; in this matrix $y_{sr}=i$ when the rater $r$ classifies subject $s$ into category $i$. The most usual thing to do is to summarize this information in a table of absolute frequencies $x_{i_1 i_2 \ldots i_R}=\#\{s \mid y_{s1}=i_1, \ldots, y_{sR}=i_R\}$ of dimension $K^R$, where the symbol # refers to "cardinal", which incorporates the frequencies of all possible combinations of classifications carried out by the $R$ raters. Therefore $x_{i_1 i_2 \ldots i_R}$ is the number of subjects classified as type $i_1$ by rater 1, type $i_2$ by rater 2, ..., or type $i_R$ by rater $R$, and $\left\{x_{i_1 i_2 \ldots i_R}\right\}$ are the observed values of a multinomial random variable of sample size $n$ and probabilities $\left\{p_{i_1 i_2 \ldots i_R}\right\}$. Let $\overline{p}_{i_1 i_2 \ldots i_R} = x_{i_1 i_2 \ldots i_R}/n$ be the observed cell proportion, $\overline{p}_i = \overline{p}_{ii\ldots i}=\#\{s \mid y_{s1}=\ldots= y_{sR}=i\}/n=x_{ii\ldots i}/n=x_i/n$ the observed proportion of agreements in category $i$ and its average value $p_i=p_{ii\ldots ii}$, $\overline{p}=\sum_{i=1}^{K}\overline{p}_i$ the observed total proportion of agreements and its average value $p=\sum_{i=1}^{K}p_i$ and, finally, $\overline{t}_{ir}=\sum_{i_1=1}^{K}\cdots\sum_{i_{r-1}=1}^{K}\sum_{i_{r+1}=1}^{K}\cdots\sum_{i_R=1}^{K}\overline{p}_{i_1\ldots i_r=i\ldots i_R}=\#\{s \mid y_{sr}=i\}/n$ the observed total proportion of responses $i$ of rater $r$ and its average value $t_{ir}=\sum_{i_1=1}^{K}\cdots\sum_{i_{r-1}=1}^{K}\sum_{i_{r+1}=1}^{K}\cdots\sum_{i_R=1}^{K}p_{i_1\ldots i_r=i\ldots i_R}$.

The $\kappa_C$ coefficient can be generalized to the case of multi-rater in several ways,

depending on how the phrase “an agreement occurs” is interpreted. If $A_0$ is the observed number of agreements, Max $A_0$ is the maximum number and E($A_0$) is the average value of $A_0$ on the assumption of independence among all the raters, then Hubert (1977) indicated that the estimate of the degree of agreement of type *kappa* is given by

$$\hat{\kappa}=\frac{A_0-\mathrm{E}\left(A_0\right)}{\mathrm{Max}\left(A_0\right)-\mathrm{E}\left(A_0\right)}=1-\frac{\mathrm{Max}\left(A_0\right)-A_0}{\mathrm{Max}\left(A_0\right)-\mathrm{E}\left(A_0\right)}=1-\frac{1-\overline{I}_o}{1-\overline{I}_e}=\frac{\overline{I}_o-\overline{I}_e}{1-\overline{I}_e} \tag{4}$$

where $\overline{I}_o$=$A_0$/Max($A_o$) and $\overline{I}_e$=E($A_0$)/Max($A_o$). Hubert (1977) makes the following interpretation "an agreement occurs if and only if all raters agree on the categorization of an object". In this case, Max ($A_o$)=$n$, $A_o$= $n\,\overline{I}_o=n\sum_{i=1}^{K}\overline{p}_i=n\overline{p}$ , E($A_o$)=$n\,\overline{I}_e$=$n\sum_{i=1}^{K}\prod_{r=1}^{R}\overline{t}_{ir}$ and Hubert's *kappa* coefficients

$$\hat{\kappa}_{HR}=\frac{\overline{I}_o-\overline{I}_e}{1-\overline{I}_e},\ \text{where}\ \overline{I}_o=\Sigma_{i=1}^{K}\overline{p}_i=\overline{p}\ \text{ and }\ \overline{I}_e=\Sigma_{i=1}^{K}\Pi_{r=1}^{R}\overline{t}_{ir}, \tag{5}$$

is an estimator of the population coefficient $\kappa_{HR}$=($I_o$–$I_e$)/(1–$I_e$), where $I_o=\Sigma_{i=1}^{K}p_i=p$ and $I_e=\Sigma_{i=1}^{K}\Pi_{r=1}^{R}t_{ir}$. Martín Andrés and Álvarez Hernández (2020) recently analysed this measure and obtained its SE.

However, the most traditional approach to understanding the phrase "an agreement occurs" is to understand the phrase "an agreement occurs if, and only if, two raters categorize an object consistently" by Fleiss (1971) and Hubert (1977) or a pairwise definition of agreement. An extension of the concept is Conger's $g$-wise *kappa* (1980), with $2\leq g\leq R$, where $g$=$R$ or $g$=2 yields the two previously mentioned Hubert definitions ($R$-wise and pairwise, respectively). However, *kappa* coefficients can vary from one author to another, depending on the definition of $I_e$. The most traditional definitions are those of the Fleiss *kappa* ($\kappa_F$) (Fleiss 1971) and that of Hubert’s *kappa* ($\kappa_{H2}$) (Hubert 1977) estimated by

$$\hat{\kappa}_F=1-\frac{nR^2-\Sigma_{s=1}^{n}\Sigma_{i=1}^{K}R_{si}^2}{nR\left(R-1\right)\left\{1-\Sigma_{i=1}^{K}R_i^2\right\}}\ \text{ and }\ \hat{\kappa}_{H2}=1-\frac{R^2-\left(\Sigma_{s=1}^{n}\Sigma_{i=1}^{K}R_{si}^2/n\right)}{R\left(R-1\right)-2\Sigma_{i=1}^{K}\Sigma_{r=1}^{R}\Sigma_{r'=r+1}^{R}\overline{t}_{ir}\overline{t}_{ir'}}, \tag{6}$$

where $R_{si}$=#$\{r \mid y_{sr}=i\}$=0, 1, …, $R$ is the number of raters that classify subject $s$ in category $i$, $R_i = R_{\bullet i}/nR$ is the total proportion of responses $i$ (any rater) and $R_{\bullet i}=\sum_{s=1}^{n} R_{si}$ . In the case of $\hat{\kappa}_{H2}$, the expression derives from the fact that $A_o$=$\sum_{s=1}^{n}\sum_{i=1}^{K} R_{si}\left(R_{si}-1\right)/2$, Max($A_0$)=$nR$($R$−1)/2 and E($A_o$)=$n\sum_{i=1}^{K}\sum_{r=1}^{R}\sum_{r'=r+1}^{R}\bar{t}_{ir}\bar{t}_{ir'}$. Warrens (2010) proved that $\hat{\kappa}_F \le \hat{\kappa}_{H2}$ and that if all the raters classify all the subjects, then $\hat{\kappa}_{H2}$ is more appropriate. It can be seen that when $R$=2 then $\hat{\kappa}_{HR}=\hat{\kappa}_{H2}=\hat{\kappa}_C$, but $\hat{\kappa}_F \neq \hat{\kappa}_C$.

The extension of the *multi-rater delta* model for $R$=2 -expressions (2)- to the case of many raters is immediate. Now

$$p_{i_1 i_2 \ldots i_R}=\delta_{i_1 i_2 \ldots i_R}\alpha_{i_1}+\left(1-\Delta\right)\prod_{r=1}^{R}\pi_{i_r r} \text{ with } \Delta=\sum_{i=1}^{K}\alpha_i \text{ and } 1=\sum_{i_r=1}^{K}\pi_{i_r r} \qquad \textbf{(7)}$$

where $i_r$=1, 2, …, $K$, and $\Delta$≤1, $\alpha_i$≤1 and 0≤$\pi_{i_r r}$≤1 are the parameters of the *multi-rater delta* model. The parameters of interest, with similar interpretations to the interpretations of case $R$=$2$, are $\Delta$ and the following extension of expression (3)

$$\mathcal{S}_i=\frac{R\alpha_i}{\sum_{r=1}^{R} t_{ir}}=\frac{R\alpha_i}{R\alpha_i+\left(1-\Delta\right)\sum_{r=1}^{R}\pi_{ir}} \qquad \textbf{(8)}$$

In the *multi-rater delta* model, note that the possibility that $\mathcal{S}_i$<0, $\alpha_i$<0 or $\Delta$<0 is also allowed because the degree of agreement can sometimes be negative.

It can be seen that the coefficient $\Delta$ can be put into the traditional *kappa* format. Since $p_i$=$\alpha_i$+(1−$\Delta$)$\Pi_{r=1}^{R}\pi_{ir}$ through expression (7), then by adding up in $i$ and working out $\Delta$ we obtain

$$\Delta=\frac{I_o-I_\pi}{1-I_\pi} \text{ where } I_\pi=\sum_{i=1}^{K}\Pi_{r=1}^{R}\pi_{ir}, \qquad \textbf{(9)}$$

so that the expected index of agreements under the *multi-rater delta* model is obtained on the basis of the probabilities $\pi_{ir}$ instead of on the basis of the probabilities $t_{ir}$ of the marginal distributions.

## 4. Estimation with the *delta* model

### *4.1. General case of more than two raters or more than two categories* ($R$>2 or $K$>2)

As can be seen in Appendix B, and because of the structure of the model, to make inferences with the *multi-rater delta* model, only some of the observed cell proportions $\bar{p}_{i_1 i_2 \ldots i_R}$ are needed: $\bar{p}_i$ (the observed proportions of agreements in category $i$) and $\bar{d}_{ir}=\bar{t}_{ir}-\bar{p}_i$ (the observed proportions of disagreements by rater $r$ in category $i$). Based on these proportions, the total proportion of agreements $\bar{p}=\sum_{i=1}^{K}\bar{p}_i$, the total proportion of disagreements $\bar{D}=\sum_{i=1}^{K}\bar{d}_{ir}$ (which is the same for all raters), and the total proportion of disagreements in category $i$ $\bar{D}_i=\sum_{r=1}^{R}\bar{d}_{ir}$ are determined. Based on the definitions, $\bar{t}_{ir}=\bar{p}_i+\bar{d}_{ir}$, $1=\bar{p}+\bar{D}$, $\bar{D}=\sum_{i=1}^{K}\bar{D}_i/R$ and $\bar{D}_i\leq(R-1)\bar{D}$. In addition, it can be seen that the values $R_i$ of $\hat{\kappa}_F$ are given by $R_i=\bar{p}_i+\bar{D}_i/R$.

Once these proportions are known, Appendix B shows that the estimators of the maximum likelihood ($\hat{\pi}_{ir},\hat{\alpha}_i,\hat{\varDelta}$ and $\hat{\mathcal{S}}_i$) of the various parameters of interest ($\pi_{ir},\alpha_i,\varDelta$ and $\mathcal{S}_i$) are as follows:

$$\hat{\pi}_{ir}=\frac{\lambda_i+\bar{d}_{ir}}{B},\ \hat{\alpha}_i=\bar{p}_i-\lambda_i,\ \hat{\varDelta}=1-B \text{ and } \hat{\mathcal{S}}_i=\frac{R\hat{\alpha}_i}{\sum_{r=1}^{R}\bar{t}_{ir}}=\frac{R\hat{\alpha}_i}{R\bar{p}_i+\bar{D}_i}=\frac{R\hat{\alpha}_i}{R\hat{\alpha}_i+\left(1-\hat{\varDelta}\right)\sum_{r=1}^{R}\hat{\pi}_{ir}} \quad \textbf{(10)}$$

where $B=1-\varDelta\geq 0$ and $\lambda_i=B\sum_{r=1}^{R}\pi_{ir}\geq 0$ are the solutions of expressions

$$B^{R-1}=\frac{\prod_{r=1}^{R}\left(\lambda_i+\bar{d}_{ir}\right)}{\lambda_i}\ (\forall i \mid \bar{d}_{ir}\neq 0,\ \forall r) \text{ under the condition } g(B)=\sum_{i=1}^{K}\lambda_i-B+\bar{D}=0 \quad \textbf{(11)}$$

with the exception that $\lambda_i=0$ when $\bar{d}_{ir}=0$ for some rater $r$. In addition, the *Supplementary Material* explicitly sets out how to proceed in order to determine the values of $B$ and $\lambda_i$, with details on how to act in the extreme cases of $B$=0 and $B$=∞. In the special case of independence among raters, that is, $\overline{p}_i = \prod_{r=1}^{R} \overline{t}_{ir} = \prod_{r=1}^{R} \left( \overline{p}_i + \overline{d}_{ir} \right)$, then $\lambda_i = \overline{p}_i$ and $B$=1 are the solutions of expressions (11); thus, $\hat{\alpha}_i = \hat{\Delta} = \hat{\kappa}_{HR}$=0. When $\overline{I}_o = \sum_{i=1}^{K} \overline{p}_i$=1, in which case $\hat{\kappa}_{HR}$=1, then $\overline{d}_{ir}$=0 ($\forall i$, $r$), $\lambda_i$=0 ($\forall i$), $g(B)$=0 implies that $B = \overline{D}$=0 and $\hat{\alpha}_i = \hat{\Delta} = \hat{\kappa}_{HR}$=1. In general, $\hat{\Delta}$ and $\hat{\kappa}_{HR}$ take similar values, except when the marginals are very unbalanced, in which case $\hat{\Delta}$ is usually superior to $\hat{\kappa}_{HR}$ (please see the examples in Section 5 and at the end of Appendix B).

Once the parameter estimates of the *multi-rater delta* model are known, the classical chi-square test of the goodness of fit of the model can be performed. Since the observed quantities and expected quantities are $n\overline{p}_{i_1 i_2 \ldots i_R}$ and $n\hat{p}_{i_1 i_2 \ldots i_R} = n\delta_{i_1 i_2 \ldots i_R} \hat{\alpha}_{i_1} + n\left(1 - \hat{\Delta}\right) \prod_{r=1}^{r=R} \hat{\pi}_{i_r r}$, respectively, then the type I $\alpha$ error test consists of comparing

$$\chi^2_{exp} = n\left[ \sum_{i_1=1}^{K} \cdots \sum_{i_R=1}^{K} \frac{\left( \overline{p}_{i_1 \ldots i_R} - \hat{p}_{i_1 \ldots i_R} \right)^2}{\hat{p}_{i_1 \ldots i_R}} \right] = n\left[ \frac{1 - \delta_{i_1 \ldots i_R}}{1 - \hat{\Delta}} \sum_{i_1=1}^{K} \cdots \sum_{i_R=1}^{K} \frac{\overline{p}^2_{i_1 \ldots i_R}}{\prod_{r=1}^{r=R} \hat{\pi}_{i_r r}} - \overline{D} \right] \quad \textbf{(12)}$$

*vs* the (1−$\alpha$)-percentile $\chi^2_{\alpha, df}$ of the chi-square distribution with $df$= ($K^R$−1)−$K$−$R$($K$−1) degrees of freedom. As shown in Appendix B, the second equality of expression (12) is attributed to the fact that the observed and expected proportions of agreements ($\overline{p}_i = \hat{p}_i$) and disagreements ($\overline{d}_{ir} = \hat{d}_{ir}$) are equal, which implies that the observed and expected marginal distributions are also equivalent ($\overline{t}_{ir} = \hat{t}_{ir}$). The value of $df$ is attributed to the fact that the multinomial distribution consists of ($K^R$−1) cells that are free to take values, from which we have to subtract $K$ parameters $\alpha_i$ estimated and $R$($K$−1) probabilities $\pi_{ir}$ estimated. Applying the classic validity rule, the previous goodness-of-fit test will be valid if none of the expected quantity $n\hat{p}_{i_1 i_2 \ldots i_R}$ is lower than 1 and no more than 20% of them are lower than or equal to 5.

When the overall degree of agreement is high and the sample size is small, many of the observed and expected frequencies will be small and the previous test will not be reliable (at least in its significant results).

The variances of $\hat{\Delta}$, $\hat{\alpha}_i$ and $\hat{\mathcal{S}}_i$ must be obtained to make inferences about the measures of agreement $\Delta$, $\alpha_i$ and $\mathcal{S}_i$ . Appendix C shows that

$$\hat{V}\left(\hat{\Delta}\right) = \frac{1-\hat{\Delta}}{n}\left\{\hat{\Delta} + \frac{\hat{X}}{(R-1)\hat{X}-1}\right\}, \tag{13}$$

$$\hat{V}\left(\hat{\alpha}_i\right) = \frac{1}{n}\left[\hat{\alpha}_i\left(1-\hat{\alpha}_i\right) + \left(1-\hat{\Delta}\right)\hat{X}_i\left\{\frac{(R-1)\hat{X}_i}{(R-1)\hat{X}-1} - 1\right\}\right] \text{ and} \tag{14}$$

$$\hat{V}\left(\hat{\mathcal{S}}_i\right) = \frac{R^2}{n\bar{N}_i^2}\left[\begin{array}{l} n\hat{V}\left(\hat{\alpha}_i\right) - \hat{\alpha}_i\left(1-\hat{\alpha}_i\right) + \hat{\alpha}_i\left(1-\hat{\mathcal{S}}_i\right)\left\{1-\frac{R-1}{R}\hat{\mathcal{S}}_i\right\} + \\ + \frac{\left(1-\hat{\Delta}\right)\hat{\mathcal{S}}_i^2}{R^2}\left\{\left(\sum_{r=1}^{R}\hat{\pi}_{ir}\right)^2 - \left(\sum_{r=1}^{R}\hat{\pi}_{ir}^2\right)\right\} \end{array}\right] \tag{15}$$

where $\hat{X}_i = \left[\sum_{r=1}^{R}\hat{\pi}_{ir}^{-1} - \left(\prod_{r=1}^{R}\hat{\pi}_{ir}\right)^{-1}\right]^{-1}$ and $\hat{X} = \sum_{i=1}^{K}\hat{X}_i$ . These estimated variances cannot be applied when any of the estimated parameters are at the boundary of the parametric space or are indeterminate. This condition occurs when $\bar{d}_{ir}$=0 or $B$=∞ because then there exists some $\hat{\pi}_{ir}$=0, as shown in *Supplementary Material*. In these cases, variances can be estimated if the calculations are carried out for the data $x_{i_1 i_2 \ldots i_R}$ increased by .5; thus, the new sample size is $n$+$K^R$/2, and the new proportions observed are $\bar{p}_{i_1 i_2 \ldots i_R} = \left(x_{i_1 i_2 \ldots i_R} + 0.5\right) / \left(n + K^R/2\right)$.

As shown in *Supplementary Material*, all results obtained by the current *multi-rater delta* model are compatible with the results of the classic *delta* model, which is specifically defined for $R$=2 and $K$>2, with the exception of some very specific results in which errors were made by Martín Andrés & Femia Marzo (2004 and 2005). When one of these two raters is a gold-standard or when the marginal distribution is fixed beforehand, the classic *delta* model is preferred since it contemplates these two situations.

*4.2. Special case with only two raters and only two categories (R=K=*2)

When only two raters and two categories exist, the problem with the *multi-rater delta* model is that there are more unknown parameters ($\alpha_1$, $\alpha_2$, $\pi_{11}$ and $\pi_{12}$) than free cells to take values (of which there are only three). In this case, the following solution by Martín Andrés & Femia Marzo (2004) can be adopted; it has been proved to provide coherent results (Martín Andrés & Femia Marzo 2004, 2005 and 2008; Ato *et al.* 2011; Shankar and Bangdiwala 2014, Giammarino *et al.* 2021). The procedure is to create a third dummy category of observed frequencies $x_{i3}=x_{3j}=0$ ($\forall i$, $j$), increase all data in the new 3×3 table by .5, estimate the parameters as executed in the previous section, and redefine the measures of agreement without considering the third dummy category. Let $\bar{p}_i$ and $\bar{d}_{ir}$ be the new observed proportions and $\alpha_i$, $\pi_{ir}$ and $\Delta$ be the parameters of the *multi-rater delta* model; all parameters refer to the new 3×3 table. The measures of agreement for the original 2×2 table are defined as $\alpha_i^* = \alpha_i/(p_{1\bullet}+p_{2\bullet})$, $\Delta^* = \alpha_1^* + \alpha_2^*$ and $\mathcal{S}_i^* = 2\alpha_i/(p_{1\bullet}+p_{2\bullet})$, for $i$=1 and 2; their estimates and estimated variances (see Appendix D) are

$$\hat{\alpha}_i^* = \frac{\hat{\alpha}_i}{1-\bar{p}_{3\bullet}},\ \hat{\mathrm{V}}\left(\hat{\alpha}_i^*\right) = \frac{1}{n\left(1-\bar{p}_{3\bullet}\right)^2}\left[\left(1-\hat{\Delta}\right)\hat{X}_i\left\{\frac{\hat{X}_i}{\hat{X}-1}-1\right\}+\left(1-\bar{p}_{3\bullet}\right)\hat{\alpha}_i^*\left(1-\hat{\alpha}_i^*\right)\right], \quad \textbf{(16)}$$

$$\hat{\Delta}^* = \hat{\alpha}_1^* + \hat{\alpha}_2^*,\ \hat{\mathrm{V}}\left(\hat{\Delta}^*\right) = \frac{1}{n\left(1-\bar{p}_{3\bullet}\right)^2}\left[\left(1-\hat{\Delta}\right)\left(1-\hat{X}_3\right)\frac{\hat{X}-\hat{X}_3}{\hat{X}-1}+\left(1-\bar{p}_{3\bullet}\right)\hat{\Delta}^*\left(1-\hat{\Delta}^*\right)\right], \quad \textbf{(17)}$$

$$\hat{\mathcal{S}}_i^* = \frac{2\hat{\alpha}_i}{\bar{N}_i},\ \hat{\mathrm{V}}\left(\hat{\mathcal{S}}_i^*\right) = \frac{4}{n\bar{N}_i^2}\left[\left(1-\hat{\Delta}\right)\hat{X}_i\left\{\frac{\hat{X}_i}{\hat{X}-1}-1\right\}+\hat{\alpha}_i\left\{1-\frac{3\hat{\alpha}_i}{\bar{N}_i}+\frac{2\hat{\alpha}_i^2}{\bar{N}_i^2}+\frac{2\hat{\pi}_{i1}\hat{\pi}_{i2}\left(1-\hat{\Delta}\right)\hat{\alpha}_i}{\bar{N}_i^2}\right\}\right], \quad \textbf{(18)}$$

where $\left(1-\bar{p}_{3\bullet}\right) = \bar{p}_1 + \bar{p}_2 + \bar{d}_{11} + \bar{d}_{21}$.

## 5. Examples

In this section, the data for the two examples described in Section 1 will be analysed,

as well as for two other examples which are a modification of the first two. As regards the data for the *R*=2 raters in Table 1(a), it has already been pointed out in Section 2 that $\hat{\kappa}_C$=.6765. With only two raters, $\kappa_F$ is not usually calculated; its estimation $\hat{\kappa}_F$=.6753 is different from the value of Cohen' *kappa*. To apply the *multi-rater delta* model the first step is to construct Table 1(b), which is the appropriate table for this model; for example $\overline{p}_1=\overline{p}_{11}$=.75, $\overline{d}_{11}=\overline{p}_{12}+\overline{p}_{13}$=.05, etc. Based on this table and the expression (11), the values *B*=.3125, $\lambda_1$=.0500, $\lambda_2$=.0048 and $\lambda_3$=0 (which is directly derived from the fact that $\overline{d}_{31}$=0) are obtained. Finally, through expressions (10), the results of Table 1(c) are obtained. According to the *multi-rater delta* model, the degree of overall agreement $\hat{\Delta}$=.6875 (which is very close to $\hat{\kappa}_C$) is the sum of the proportion of agreements not due to chance obtained in each of the three classes ($\hat{\alpha}_1+\hat{\alpha}_2+\hat{\alpha}_3$=.5500+.0375+.1000), but the degree of agreement in each class is given by the consistencies $\hat{\mathcal{S}}_1=.6875$, $\hat{\mathcal{S}}_2=.5000$ and $\hat{\mathcal{S}}_3=.8000$; hence class 3 is the one with the best degree of agreement between the two raters, while class 2 has the worst degree of agreement. These values are also quite close to those given by *kappa* when the data in Table 1(a) are collapsed in each of the three classes: $\hat{\kappa}_{C(1)}$=.6875, $\hat{\kappa}_{C(2)}$=.5000 and $\hat{\kappa}_{C(3)}$=.7727. According to Martín Andrés & Femia Marzo (2005) this similarity of results occurs when the two marginal distributions are homogenous (as in the example) and they are not excessively unbalanced. As regard the SE, they have all been obtained based on the data in Table 1(a) incremented by .5, because $\hat{\pi}_{31}$=0 (which is due to the fact that $\lambda_3$=0). In reality, the first step should be to verify the validity of the model; by applying the expression (12) to the data in Table 1(a), we obtain $\chi^2_{exp}$=0 (*df*=1) so that the *multi-rater delta* model is a perfect fit.

In the previous example the difference between $\hat{\kappa}_C$=.6765 and $\hat{\Delta}$=.6875 is small, but the situation changes when the marginals have even more imbalance. By modifying Table 1(a) as in Martín Andrés & Femia Marzo (2004, 2005), in which case the data by rows are 92/0/0 (row 1), 2/1/1 (row 2) and 2/1/1 (row 3), one obtains the values $\hat{\kappa}_C$=.4792 and $\hat{\Delta}$=.9200, which are now very different. The reason for this discrepancy is that the coefficient *kappa* does not take into account the information given by the marginal distributions, unlike coefficient *delta*. Note that in this example both raters are in agreement on classifying almost all the subjects into the psychotic type (92% and 96% respectively), a fact that *delta* takes into consideration but not *kappa*. In this example, once again $\chi^2_{exp} = 0$ ($df$=1).

As regards the data for the $R$=3 raters in Table 2(a), in order to apply the *multi-rater delta* model Table 2(b) must first be constructed; for example, $n\,\overline{p}_1$=56 and $n\,\overline{d}_{11}$= 1+0+5+3+ 0+0+0+1=10. Based on this table, and the expression (11), the values $B$=0.4504, $\lambda_1$= 0.0104, $\lambda_2$=0.0774 and $\lambda_3$=0.0030 are obtained. Finally, the results in Table 2(c) are obtained from expressions (10). Note that the total agreement is 54.96% and the degrees of agreement in categories 1, 2 and 3 (consistencies) are 70.40%, 24.62%, and 63.06%, respectively. This finding indicates that category 2 has a very low consistency (24.62%), so the three raters should make efforts to homogenize their classification criteria especially in this category. These values are no longer so close to those given by Hubert's *kappa* when the data in Table 2(a) are collapsed in each of the three classes $-\hat{\kappa}_{HR(1)}$=63.62%, $\hat{\kappa}_{HR(2)}$=42.70% and $\hat{\kappa}_{HR(3)}$=60.81%−, especially in the case of class 2; the reason is that now the marginal distributions of the three raters are not homogenous, due in particular to rater 2. The raw values of these four parameters, or values uncorrected for chance, are 100/164=60.98% for the total degree of agreement and 3×56/232=72.41%, 3×20/148=40.54%, and 3×24/112= 64.29% for consistencies in categories 1, 2, and 3, respectively. The estimated total degree of agreement is .5496±.0462, which indicates that the real total degree of agreement is $\Delta\geq$

.5496−1.645×.0462=.4736 for 95% confidence. Finally, by applying the expression (12) to the data in Table 2(a), we obtain $\chi^2_{exp}$ = 155.41 (*df*=17); thus, the test is significant for a Type I error of $\alpha$=5%, and the *multi-rater delta* model does not fit the data. However, this result is not reliable because 7 (21) expected amounts less than one (less than or equal to five) exist, that is, 25.9% (77.8%); this is a normal occurrence when $R$>2.

Determining the different *kappa* measures of agreement requires summarizing the data of Table 2(a) in a different way. To calculate $\hat{\kappa}_{HR}$ one needs the data in Table 2(d), which are obtained from Table 2(b); based on this and the expression (5), $\hat{\kappa}_{HR}$=.5471. To calculate $\hat{\kappa}_F$ one needs the data in Table 2(e), which are obtained from Table 2(a); for example, the number 26 in Table 2(e), which is the number of subjects classified in category 1 by only one of the three raters, is obtained by summing all the frequencies $x_{i_1 i_2 i_3}$ of Table 2(a), in which on only one occasion $i_r$=1, that is, 26=3+1+2+1+14+1+2+2. Based on Table 2(e), $\sum_{s=1}^{164}\sum_{i=1}^{3} R_{si}^2$ = $72\times1^2+60\times2^2+100\times3^2$=1,212, $\sum_{i=1}^{3} R_i^2$ =$(232^2+148^2+112^2)/(164\times3)^2$=.3647 and, through the expression (6), $\hat{\kappa}_F$ =1−[164×9−1,212]/[164×3×2×(1−.3647)]=.5777. Finally, to calculate $\hat{\kappa}_{H2}$ the data in Tables 2(d) and (e) is needed; from the expression (5), $\hat{\kappa}_{H2}$=.5809. Table 2(f) summarizes the numerical results produced by applying the different measures of degree of agreement to the data in Table 2(a), where the raw degree of agreement is $\sum_{i=1}^{3} \overline{p}_i$ =(56+20+24)/164=.6098.

In the previous example, the differences between the various measures of the overall degree of agreement corrected for chance are small; however, the situation changes when the marginals are unbalanced. This is the case with the data in Table 3(a), which, as they are a modification of the data in Table 2(a), yield very unbalanced marginals. For example, the proportions of responses by rater 1 in categories 1, 2 and 3 are 115/164, 27/164, and 22/164, respectively. The results in Table 3(e) indicate that the *multi-rater delta* degree of agreement (70.8%) is slightly lower than the degree of raw agreement (74.4%), but significantly higher than the *kappa*-based agreement (57.4%, 55.5% and 55.4%), which is attributed to the fact that the *kappa* coefficients are influenced by the marginal distributions (because the *kappa* coefficients do not take into account the information provided by the marginal about the degree of agreement). Now $\chi^2_{exp}$=19.83 (*df*=17) is not significant for $\alpha$=5%, even though 9 (24) expected amounts less than one (less than or equal to five) exist. At the end of Appendix B, a theoretical study is carried out on the value of the difference $\hat{\Delta}-\hat{\kappa}_{HR}$.

## 6. Conclusions

In this paper, we have looked at the evaluation of multi-rater agreement in the case of nominal categories. This issue is important in medicine, psychometry and whenever the intention has been to measure the degree of agreement among *R* raters who classify *n* subjects within *K* categories.

When only two raters (*R*=2) exist, it is very common to use Cohen's *kappa* coefficient (1960) and, occasionally, the *delta* coefficient (Martín Andrés and Femia Marzo 2004). The classic *delta* coefficient has three advantages over the *kappa* coefficient. The first advantage is that the *delta* coefficient is not affected by imbalance in the marginal distributions of each rater. The second advantage is that the *delta* model from which this coefficient is derived has no difficulty in evaluating the degree of agreement in each category (Martín Andrés & Femia Marzo, 2005). The opposite occurs in the case of the *kappa* coefficient (Brennan & Prediger 1981; Agresti *et al.*, 1995; Guggenmoos-Holzmann & Vonk, 1998). Finally, the *delta* coefficient is easier to interpret than the *kappa* coefficient; both refer to the proportion of replies that are concordant not by chance, but population references are different: the first one is the total number of responses (*n*) and the second one is the maximum number possible of non-random agreements ($1-\overline{I}_e$). In the first case, the coefficient is interprets as "$\Delta\times 100$% of the responses are concordant not by chance"; in the second case "the two raters agree on $\kappa_C$×100% of the maximum number possible of non-random agreements".

When many raters ($R$≥2) exist, several *kappa*-based coefficients can be employed, such as Fleiss' *kappa* (1971), Hubert's *kappa* (1977), and Conger's *kappa* (1980). However, Fleiss' *kappa* does not coincide with Cohen's *kappa* when $R$=2. The limitations of all *multi-rater kappa* coefficients are identical to the limitations of the Cohen's *kappa* coefficient (Marasini *et al.*, 2016; Conger, 2017). In this article, the classic *delta* model has been modified to ensure that it is valid in the multi-rater case; thus, the *multi-rater delta* model is obtained. This model provides results that are compatible with the results of the classic *delta* model, so it has the same advantages over the *multi-rater kappa* coefficients. In the first place, the degree of total agreement of the *multi-rater delta* model (parameter $\Delta$) is not affected by the marginal distributions of the raters, unlike the *multi-rater kappa* coefficients. Second, the *multi-rater delta* model allows the degree of agreement in each class to be measured through the concept of consistency (coefficients $\mathcal{S}_i$). However, *multi-rater kappa* coefficients usually measure the degree of agreement in each class by collapsing the data in this class, this means that the collapsed *kappa* coefficients attempt to measure simultaneously the degree of agreement in this class, the degree of agreement in all the remaining classes as a single class and the whole degree of agreement in the collapsed table. It is not possible to reconcile these three objectives simultaneously. Lastly, the interpretation of the *multi-rater delta* coefficient is also simpler than the interpretation of *multi-rater kappa* coefficients. A free program (Multi_Rater_ Delta.exe) may be downloaded to run this model at http://www.ugr.es/local/bioest/software (Section "Agreement Among Raters").

**Acknowledgements**

The authors wish to thank Dr. Pedro Femia Marzo for his invaluable assistance in various aspects of this article.

**Disclosure statement**

No potential conflict of interest was reported by the authors.

## APPENDICES

### A. Relationships between the parameters of two (classic and new) *delta* models for *R*=2

Under the *delta* model given by expression (1), Martín Andrés & Femia Marzo (2004, 2005) defined the following three parameters for the case of any two raters

$$\mathcal{A}_i = p_{i\bullet}\Delta_i,\ \Delta=\Sigma\,\mathcal{A}_i \text{ and } \mathcal{S}_i = 2p_{i\bullet}\Delta_i/(p_{i\bullet}+p_{\bullet i}) \quad \textbf{(A1)}$$

where $\mathcal{A}_i$ is the proportion of the agreements in category $i$ that do not occur by chance; $\Delta$ is the overall proportion of agreements that are not due to chance (that is, the overall degree of agreement); and $\mathcal{S}_i$ is the *consistency* in category $i$, which measures the degree of agreement in category $i$. Due to expressions (1) and (A1),

$$p_{ij}=\delta_{ij}\alpha_i+(p_{i\bullet}-\alpha_i)\pi_j \text{ with } \alpha_i=p_{i\bullet}\Delta_i=\mathcal{A}_i \quad \textbf{(A2)}$$

If $\Delta=\sum_{i=1}^{K}\alpha_i=\sum_{i=1}^{K}p_{i\bullet}\Delta_i$, from expression (A2), $p_{\bullet j}=\sum_{i=1}^{K}p_{ij}=p_{jj}+\sum_{i=1,i\neq j}^{K}p_{ij}=\alpha_j+(p_{j\bullet}-\alpha_j)\pi_j+\pi_j\sum_{i=1,i\neq j}^{K}\left(p_{i\bullet}-\alpha_i\right)=\alpha_j+\pi_j\sum_{i=1}^{K}\left(p_{i\bullet}-\alpha_i\right)=\alpha_j+(1-\sum_{i=1}^{K}\alpha_i)\pi_j=\alpha_j+(1-\Delta)\pi_j$.

The *delta* model of expression (1) is expressed "in rows", that is, taking rater 1 as the reference. If the model is "in columns", that is, taking rater 2 as the reference, it will depend on new parameters $\tilde{\Delta}_j$ and $\tilde{\pi}_j$, similar to $\Delta_i$ and $\pi_i$ as previously defined. Following the reasoning in the previous paragraph, $p_{ij}=\delta_{ij}\tilde{\alpha}_j+\left(p_{\bullet j}-\tilde{\alpha}_j\right)\tilde{\pi}_i$, with $\tilde{\alpha}_j=p_{\bullet j}\tilde{\Delta}_j$, $p_{i\bullet}=\tilde{\alpha}_i+\left(1-\tilde{\Delta}\right)\tilde{\pi}_i$ and $\tilde{\Delta}=\sum_{j=1}^{K}\tilde{\alpha}_j$. Since $p_{ij}$ must take the same value in both models, $p_{ij}=(p_{i\bullet}-\alpha_i)\pi_j=\left(p_{\bullet j}-\tilde{\alpha}_j\right)\tilde{\pi}_i$ for $i\neq j$, that is, $(p_{i\bullet}-\alpha_i)/\tilde{\pi}_i=\left(p_{\bullet j}-\tilde{\alpha}_j\right)/\pi_j$ ($\forall i\neq j$). Assuming that $K>2$, since the case of $K=2$ requires special treatment, the previous equality takes the constant value $\gamma$, and so

$$p_{i\bullet}-\alpha_i = \gamma\tilde{\pi}_i \text{ and } p_{\bullet j}-\tilde{\alpha}_j=\gamma\pi_j \qquad \textbf{(A3)}$$

By adding the values in $i$ or $j$, we obtain $1-\sum_{i=1}^{K}\alpha_i=1-\sum_{i=1}^{K}\tilde{\alpha}_i=\gamma$, and $\Delta=\tilde{\Delta}$. In the case of $i=j$, $p_{ii}=\alpha_i+(p_{i\bullet}-\alpha_i)\pi_i=\tilde{\alpha}_i+\left(p_{\bullet i}-\tilde{\alpha}_i\right)\tilde{\pi}_i$, so $\alpha_i+\gamma\pi_i\tilde{\pi}_i=\tilde{\alpha}_i+\gamma\tilde{\pi}_i\pi_i$ and $\alpha_i=\tilde{\alpha}_i$. Substituting expression (A3) into (A2) and considering that $\gamma=1-\Delta$, we obtain $p_{ij}=\delta_{ij}\alpha_i+(1-\Delta)\tilde{\pi}_i\pi_j$, which is the *multi-rater delta* model of expression (2). The result is that it is irrelevant whether the *delta* model is expressed "in rows" or "in columns" because the parameters of expression (A1) are equivalent in both cases.

The values obtained in the previous paragraphs yield the parameters of the *delta* model defined "in columns" according to the parameters of the *delta* model defined "in rows":

$$p_{\bullet i}=p_{i\bullet}\Delta_i+(1-\Delta)\pi_i,\ \tilde{\Delta}_i=\frac{p_{i\bullet}\Delta_i}{p_{i\bullet}\Delta_i+(1-\Delta)\pi_i} \text{ and } \tilde{\pi}_i=\frac{p_{i\bullet}(1-\Delta_i)}{1-\Delta} \qquad \textbf{(A4)}$$

The first equality stems from the final statement in the first paragraph; the second equality from the fact that $\alpha_i=p_{i\bullet}\Delta_i=\tilde{\alpha}_i=p_{\bullet i}\tilde{\Delta}_i$; and the third equality from the fact that $\tilde{\pi}_i=(p_{i\bullet}-\alpha_i)/\gamma=(p_{i\bullet}-\tilde{\alpha}_i)/\gamma=p_{i\bullet}(1-\Delta_i)/(1-\Delta)$. Based on all the foregoing, the following relationships between the parameters of the new *delta* model (Section 2.3) and of the classic *delta* model (Section 2.2), as defined “in rows” can also be deduced.

$$\alpha_i=p_{i\bullet}\Delta_i,\quad \Delta=\sum_{i=1}^{K}p_{i\bullet}\Delta_i,\quad \pi_{i2}=\pi_i,\ \pi_{i1}=p_{i\bullet}(1-\Delta_i)/(1-\Delta),\qquad p_{i\bullet}=\alpha_i+(1-\Delta)\pi_{i1} \qquad \textbf{(A5)}$$

**B. Maximum likelihood estimators of the *multi-rater delta* model when *R*>2 or *K*>2**

To simplify this explanation, in the following proofs will be provided when $R=3$; these can be extended directly to any other case, except in the case of certain aspects that will be specified as they arise. In addition, now the *multi-rater delta* model of expression (7) is simplified to $p_{ijh}=\delta_{ijh}\alpha_i+(1-\Delta)\pi_{i1}\pi_{j2}\pi_{h3}$ with $i, j, h=1, 2, \ldots, K$ and $\Delta=\sum_{i=1}^{K}\alpha_i$. If $\overline{p}_{ijh}=x_{ijh}/n$ are the observed proportions, then, except for one constant the logarithm of the likelihood is $L=$ $\sum_{i=1}^{K}\sum_{j=1}^{K}\sum_{h=1}^{K}\overline{p}_{ijh}\log p_{ijh}=\sum_{i=1}^{K}\overline{p}_{iii}\log p_{iii}+\sum_{i=1}^{K}\sum_{j=1}^{K}\sum_{h=1}^{K}\overline{p}_{ijh}\,(1-\delta_{ijh})\log p_{ijh}=\sum_{i=1}^{K}\overline{p}_{iii}\log p_{iii}+$ $\sum_{i=1}^{K}\sum_{j=1}^{K}\sum_{h=1}^{K}\overline{p}_{ijh}\left(1-\delta_{ijh}\right)\left\{\log\left(1-\mathit{\Delta}\right)+\log\pi_{i1}+\log\pi_{j2}+\log\pi_{h3}\right\}$. By working it out we obtain $L=\sum_{i=1}^{K}\overline{p}_{iii}\log p_{iii}+(1-\overline{p})\log\left(1-\mathit{\Delta}\right)+\sum_{i=1}^{K}\sum_{r=1}^{3}\left(\log\pi_{ir}\right)\overline{d}_{ir}$, where $\overline{p}=\sum_{i=1}^{K}\overline{p}_{iii}$ is the total proportion of agreements, $r$=1, 2, 3 and $\overline{d}_{ir}$ is the proportion of disagreements in category $i$ of rater $r$, that is, $\overline{d}_{i1}=\overline{p}_{i\bullet\bullet}-\overline{p}_{iii}$, $\overline{d}_{i2}=\overline{p}_{\bullet i\bullet}-\overline{p}_{iii}$ and $\overline{d}_{i3}=\overline{p}_{\bullet\bullet i}-\overline{p}_{iii}$.

Bearing in mind that $p_{iii}=\alpha_i+(1-\mathit{\Delta})\pi_{i1}\pi_{i2}\pi_{i3}$, the following derivatives are obtained

$$\frac{\partial p_{iii}}{\partial\alpha_t}=\delta_{it}-\frac{p_{iii}-\alpha_i}{1-\mathit{\Delta}},\quad \frac{\partial p_{iii}}{\partial\pi_{tr}}=\delta_{it}\frac{p_{iii}-\alpha_i}{\pi_{ir}},$$

and the derivatives of $L$ are

$$\frac{dL}{d\alpha_i}=\frac{\overline{p}_{iii}}{p_{iii}}-\frac{A}{1-\mathit{\Delta}}\ \text{with}\ A=1-\sum_{i=1}^{K}\frac{\overline{p}_{iii}}{p_{iii}}\alpha_i,\quad \frac{\partial L}{\partial\pi_{ir}}=\frac{1}{\pi_{ir}}\left\{\overline{d}_{ir}+\overline{p}_{iii}-\frac{\overline{p}_{iii}}{p_{iii}}\alpha_i\right\}=A_{ir} \qquad \textbf{(B1)}$$

Since the maximum likelihood estimates of $\alpha_i$ must verify that $dL/d\alpha_i$=0, then $\overline{p}_{iii}/p_{iii}=A/(1-\mathit{\Delta})$ through the first expression of (B1). By substituting in the definition of $A$, $A=1-\mathit{\Delta}$, $\overline{p}_{iii}/p_{iii}=1$ and $\overline{p}_{iii}=p_{iii}$. The last equality indicates that if the maximum likelihood estimates $\hat{\alpha}_i$ and $\hat{\pi}_{ir}$ of $\alpha_i$ and $\pi_{ir}$, respectively, were known, the maximum likelihood estimator for $p_{iii}$ would be $\hat{p}_{iii}=\hat{\alpha}_i+\left(1-\hat{\mathit{\Delta}}\right)\hat{\pi}_{i1}\hat{\pi}_{i2}\hat{\pi}_{i3}=\overline{p}_{iii}$. Thus

$$\hat{\alpha}_i=\overline{p}_{iii}-\left(1-\hat{\mathit{\Delta}}\right)\hat{\pi}_{i1}\hat{\pi}_{i2}\hat{\pi}_{i3}\ \text{and}\ \hat{p}_{iii}=\overline{p}_{iii}, \qquad \textbf{(B2)}$$

where $\hat{\mathit{\Delta}}=\sum_{i=1}^{K}\hat{\alpha}_i$ is the maximum likelihood estimator of $\mathit{\Delta}$. Note that $\hat{\alpha}_i=\overline{p}_{iii}$ if $\exists r \mid \hat{\pi}_{ir}=0$.

Substituting $\overline{p}_{iii}=p_{iii}$ in the second expression of (B1), $\partial L/\partial\pi_{ir}=\left(\overline{d}_{ir}+\overline{p}_{iii}-\alpha_i\right)/\pi_{ir}=A_{ir}$. If the maximum likelihood estimators of $\pi_{ir}$ have to verify that $dL/d\pi_{ir}$=0 and $\sum_{i=1}^{K}\pi_{ir}=1$, then $\partial L/\partial\pi_{ir}=\partial L/\partial\pi_{jr}$, $A_{ir}=A_{jr}$ $(\forall i,\ j)$, $A_{ir}=A_r$ and $\sum_{i=1}^{K}A_r\pi_{ir}=A_r=1-\mathit{\Delta}$. Therefore, $A_r=A=1-\mathit{\Delta}$.

Given that $\bar{p}_{iii}-\alpha_i=p_{iii}-\alpha_i=(1-\Delta)\pi_{i1}\pi_{i2}\pi_{i3}$, from the last expression (B1), we obtain

$$\frac{\partial L}{\partial \pi_{ir}}=\frac{1}{\pi_{ir}}\left\{\bar{d}_{ir}+(1-\Delta)\pi_{i1}\pi_{i2}\pi_{i3}\right\}=1-\Delta \quad \text{or} \quad (1-\Delta)\pi_{ir}=\bar{d}_{ir}+(1-\Delta)\pi_{i1}\pi_{i2}\pi_{i3}. \qquad \textbf{(B3)}$$

These expressions yield the maximum likelihood estimators $\hat{\Delta}$ and $\hat{\pi}_{ir}$. Once the estimators are obtained, $\hat{d}_{i1}=\hat{p}_{i\bullet\bullet}-\hat{p}_{iii}=(1-\hat{\Delta})\hat{\pi}_{i1}-(1-\hat{\Delta})\hat{\pi}_{i1}\hat{\pi}_{i2}\hat{\pi}_{i3}$; thus, $\hat{d}_{i1}=\bar{d}_{i1}$ from the last equality of (B3) and $\hat{p}_{i\bullet\bullet}=\bar{p}_{i\bullet\bullet}$. In general, $\hat{d}_{ir}=\bar{d}_{ir}$ and $\bar{t}_{ir}=\hat{t}_{ir}$.

The estimators of $\Delta$ and $\pi_{ir}$ can be obtained based on the expressions (B3). If $\bar{d}_{i3}=0$, then $\partial L/\partial\pi_{i3}=(1-\Delta)\pi_{i1}\pi_{i2}\leq 1-\Delta$, $\mathrm{d}L/\mathrm{d}\pi_{i3}\leq 0$, and $\hat{\pi}_{i3}=0$; generally if $\bar{d}_{ir}=0$, then $\hat{\pi}_{ir}=0$. Let $\lambda_i=(1-\Delta)\pi_{i1}\pi_{i2}\pi_{i3}$; the last equality in expression (B3) indicates that $\pi_{ir}=(\bar{d}_{ir}+\lambda_i)/(1-\Delta)$, and therefore, $\lambda_i=(\bar{d}_{i1}+\lambda_i)(\bar{d}_{i2}+\lambda_i)(\bar{d}_{i3}+\lambda_i)/(1-\Delta)^2$ $(\forall i)$. If $B=1-\Delta$, the following expressions must be solved for $\lambda_i$:

$$B^2=\frac{\prod_{r=1}^{3}\left(\lambda_i+\bar{d}_{ir}\right)}{\lambda_i} \ (\forall i \mid \bar{d}_{ir}\neq 0,\ \forall r) \text{ under the condition } g(B)=\sum_{i=1}^{K}\lambda_i-B+\bar{D}=0, \quad \textbf{(B4)}$$

where $\lambda_i=0$ if $\exists r \mid \bar{d}_{ir}=0$. The condition $g(B)=0$ is due to the fact that since $\pi_{ir}=(\bar{d}_{ir}+\lambda_i)/B$ and $\sum_{i=1}^{K}\pi_{ir}=1$, then $B\sum_{i=1}^{K}\pi_{ir}=\bar{D}+\sum_{i=1}^{K}\lambda_i=B$. Expression (B4) is generalized in expression (11); in *Supplementary Material* it is shown how to obtain the solutions to this expression. Once the values of $B$ and $\lambda_i$ are determined, taking into account that $\pi_{ir}=(\bar{d}_{ir}+\lambda_i)/B$, the first equality of (B2), the definition of (8), and the fact that the observed and expected marginal distributions are equal, the following estimators are generalized in expression (10):

$$\hat{\pi}_{ir}=\frac{\lambda_i+\bar{d}_{ir}}{B},\ \hat{\alpha}_i=\bar{p}_{iii}-\lambda_i,\ \hat{\Delta}=1-B \text{ and } \hat{\mathcal{S}}_i=\frac{3\hat{\alpha}_i}{\bar{p}_{i\bullet\bullet}+\bar{p}_{\bullet i\bullet}+\bar{p}_{\bullet\bullet i}}=\frac{3\hat{\alpha}_i}{3\bar{p}_i+\bar{D}_i}, \qquad \textbf{(B5)}$$

Note that by substituting $\lambda_i+\bar{d}_{ir}=B\hat{\pi}_{ir}$ in the first expression (11) and by calculating we obtain

$$B = \frac{\bar{d}_{ir}}{\hat{\pi}_{ir} - \prod_{r'=1}^{R} \hat{\pi}_{ir'}} \quad (\forall i, r), \tag{B6}$$

so that the values of $\hat{\pi}_{ir}$ depend solely on the values of $\bar{d}_{ir}$ and not on the values of $\bar{t}_{ir}$.

By reason of expression (9)

$$\hat{\Delta} = \frac{\bar{I}_o - \hat{I}_\pi}{1 - \hat{I}_\pi}, \text{ where } \hat{I}_\pi = \sum_{i=1}^{K} \Pi_{r=1}^{R} \hat{\pi}_{ir},$$

and through the expression (5),

$$\hat{\Delta} - \hat{\kappa}_{HR} = \frac{\left(1 - \hat{\Delta}\right)\left(1 - \hat{\kappa}_{HR}\right)}{\left(1 - \bar{I}_e\right)} \left(\bar{I}_e - \hat{I}_\pi\right), \tag{B7}$$

so that $\hat{\Delta} - \hat{\kappa}_{HR}$ is proportional to $\bar{I}_e - \hat{I}_\pi$. The maximum (1) and minimum (0) values of $\bar{I}_e$ are reached respectively when $\bar{t}_{i1} = \bar{t}_{i2} = ... = \bar{t}_{iR} = 1$ in any $i$ (in which case all the remaining $\bar{t}_{jr}$=0) or when $\{\bar{t}_{i1}, \bar{t}_{i2}, ..., \bar{t}_{iR}\}$ in all the sets, one of the terms has a value of 1 and the rest have a value of 0. In the same way for $\hat{I}_\pi$, but now based on the probabilities $\hat{\pi}_{ir}$. In most of the practical situations the values of $\bar{t}_{ir}$ and $\hat{\pi}_{ir}$ are not close to these extreme cases, so that the value of $\bar{I}_e - \hat{I}_\pi$ is not excessive and the value of $\hat{\Delta} - \hat{\kappa}_{HR}$ will not be either. Usually, the marginal distributions $\bar{t}_{ir}$ are homogenous; if they were totally homogenous then $\bar{t}_{ir} = \bar{t}_i$ ($\forall r$), that is $\bar{d}_{ir} = \bar{d}_i$ ($\forall r$) and, through expression (B6), $\hat{\pi}_{ir} = \hat{\pi}_i$ ($\forall r$), where

$$B = \frac{\bar{d}_i}{\hat{\pi}_i - \hat{\pi}_i^R} \quad (\forall i) \text{ and } \sum_{i=1}^{K} \bar{t}_i = \sum_{i=1}^{K} \hat{\pi}_i = 1. \tag{B8}$$

The result of this is that $\bar{I}_e = \sum_{i=1}^{K} \bar{t}_i^R$ and $\hat{I}_\pi = \sum_{i=1}^{K} \hat{\pi}_i^R$. In these cases, the minimum (or maximum) value of $\bar{I}_e$ is $1/K^{R-1}$ (or 1), which is reached when $\bar{t}_i$=1/$K$ (or when all the $\bar{t}_i$ have a value of 0, except one which has a value of 1); it is similar with $\hat{I}_\pi$, but now based on the probabilities $\hat{\pi}_i$. When the marginal distributions of the proportions of responses are totally balanced, $\bar{t}_i = 1/K$ ($\forall i$), $\bar{I}_e$ takes the minimum value and $\hat{\Delta} \leq \hat{\kappa}_{HR}$ through expression (B7). When the marginal distributions of the proportions of disagreements are totally balanced, $\bar{d}_i = \bar{D}/K$ ($\forall i$), $B = \bar{D}/K\left\{\hat{\pi}_i - \hat{\pi}_i^R\right\}$ ($\forall i$), $\hat{\pi}_i = 1/K$ ($\forall i$), $\hat{I}_\pi$ takes the minimum value and $\hat{\Delta} \geq \hat{\kappa}_{HR}$ through expression (B7). However, as was previously pointed out, the important differences between $\hat{\Delta}$ and $\hat{\kappa}_{HR}$ occur when $\bar{I}_e$ approaches its maximum value. If $\bar{t}_1$ is very large, the rest of the values of $\bar{t}_i$ will be small and $\bar{I}_e$ will be close to 1; but that large marginal unbalance does not imply that the values $\bar{d}_i$ are also very unbalanced, so that, through expression (B8), the values of $\hat{\pi}_i$ will not be very unbalanced, $\hat{I}_\pi$ will not be close to its maximum value (which is 1), $\hat{I}_\pi$ will be much smaller that $\bar{I}_e$ and, finally, $\hat{\Delta}$ will be appreciably larger than $\hat{\kappa}_{HR}$.

## C. Variances of estimates when *R*>2 or *K*>2

Agresti (2013) specifies the multivariate delta method in the case of a multinomial. If a function $f=f(\boldsymbol{p_{ijh}})$ is estimated by $\hat{f} = f\left(\hat{\boldsymbol{p}}_{\boldsymbol{ijh}}\right)$, with $\boldsymbol{p_{ijh}}=\{p_{ijh}\}$, $\hat{\boldsymbol{p}}_{\boldsymbol{ijh}} = \left\{\hat{p}_{ijh}\right\}$ and $\hat{p}_{ijh}$ the maximum likelihood estimates of $p_{ijh}$, then if $f_{ijh} = \partial f/\partial p_{ijh}$, the variance of $\hat{f}$ is

$$V\left(\hat{f}\right) = \left[\sum\nolimits_{i=1}^{K}\sum\nolimits_{j=1}^{K}\sum\nolimits_{h=1}^{K} f_{ijh}^2 p_{ijh} - \left(\sum\nolimits_{i=1}^{K}\sum\nolimits_{j=1}^{K}\sum\nolimits_{h=1}^{K} f_{ijh}\right)^2\right]\Big/ n. \qquad \textbf{(C1)}$$

A similar notation to the one used for the observed proportions ($\bar{p}_{ijh}, \bar{d}_{ir}, \bar{p}_i, \bar{t}_{ir},$) will be applied for the different probabilities ($p_{ijh}$, $d_{ir}$, $p_i$, and $t_{ir}$). Any condition that an estimator must verify with respect to the observed proportions should also occur with respect to the probabilities. Since $\lambda_i/(\lambda_i+d_{ir})=B\pi_{i1}\pi_{i2}\pi_{i3}/B\pi_{ir}$, then

$$\frac{\lambda_i}{\lambda_i + d_{ir}} = \frac{P_i}{\pi_{ir}} \text{ and } \sum\nolimits_{r=1}^{R}\frac{\lambda_i}{\lambda_i + d_{ir}} = P_i I_i \text{ with } P_i = \prod\nolimits_{r=1}^{R}\pi_{ir} \text{ and } I_i = \sum\nolimits_{r=1}^{R}\pi_{ir}^{-1}. \qquad \textbf{(C2)}$$

Let $X_i=P_i/(P_iI_i-1)=1/(I_i-P_i^{-1})$ and $X=\sum\nolimits_{i=1}^{K} X_i$.

When $\hat{f}=1-\hat{\Delta}$, we have $f$=$B$ and $V(B)=V(\hat{\Delta})$, with $B$=1−$\Delta$ given by expression (11), which is expressed in terms of probabilities. Because $g(B)$=0 and $\sum_{i=1}^{K}\sum_{j=1}^{K}\sum_{h=1}^{K}p_{ijh}=1$, 0=d$g$/d$p_{ijh}$=$\partial g/\partial p_{ijh}-\partial g/\partial p_{KKK}=\partial g/\partial p_{ijh}$ since $p_{KKK}$ does not intervene in $g$. Therefore, 0= $\partial g/\partial p_{ijh}=(\partial g/\partial B)(\partial B/\partial p_{ijh})+(\partial g/\partial p_{ijh})=-\Delta'_{ijh}[\Sigma_t\lambda'_{t(B)}-1]+[\lambda'_{t(ijh)}+D'_{ijh}]$, with $\Delta'_{ijh}=\partial\Delta/\partial p_{ijh}$, $\lambda'_{t(B)}=\partial\lambda_t/\partial B$, $\lambda'_{t(ijh)}=\partial\lambda_t/\partial p_{ijh}$ and $D'_{ijh}=\partial D/\partial p_{ijh}$. Therefore,

$$\Delta'_{ijh}=\frac{\sum_{t=1}^{K}\lambda'_{t(ijh)}+D'_{ijh}}{\sum_{t=1}^{K}\lambda'_{t(B)}-1}. \qquad \textbf{(C3)}$$

For the first equality of (11), 2log $B$=$\sum_{r=1}^{R}\log(\lambda_t+d_{tr})$−log $\lambda_t$; deriving this expression, we obtain $\lambda'_{t(B)}$=2$X_t$, and the denominator of expression (C3) will be 2$X$−1. To calculate the numerator, we need $D'_{ijh}=1-\delta_{ijh}$ and $\lambda'_{t(ijh)}$. If $i$=$j$=$h$, then $\lambda'_{t(ijh)}$=0 since the first equality of (11) does not depend on $p_{iii}$. Deriving this equality with respect to $p_{ijh}$, we obtain $\lambda'_{t(ijh)}[1-P_tI_t]=\sum_{r=1}^{R}(P_t/\pi_{tr})(\partial d_{tr}/\partial p_{ijh})$ and $\lambda'_{t(ijh)}=-(1-\delta_{ijh})X_t\{(\delta_{ti}/\pi_{t1})+(\delta_{tj}/\pi_{t2})+(\delta_{th}/\pi_{t3})\}$; therefore, $\sum_{t=1}^{K}\lambda'_{t(ijh)}=\lambda'_{i(ijh)}+\lambda'_{j(ijh)}+\lambda'_{h(ijh)}=-(1-\delta_{ijh})[1-\{(X_i/\pi_{i1})+(X_j/\pi_{j2})+(X_h/\pi_{h3})\}]$. Replacing everything in expression (C3),

$$\Delta'_{ijh}=f_{ijh}=\frac{1-\delta_{ijh}}{2X-1}\left[1-\left\{\frac{X_i}{\pi_{i1}}+\frac{X_j}{\pi_{j2}}+\frac{X_h}{\pi_{h3}}\right\}\right], \qquad \textbf{(C4)}$$

where the number 2 will be ($R$−1). To apply expression (C1), we must consider that $(1-\delta_{ijh})p_{ijh}=(1-\delta_{ijh})(\delta_{ijh}\alpha_i+B\pi_{i1}\pi_{j2}\pi_{h3})=(1-\delta_{ijh})B\pi_{i1}\pi_{j2}\pi_{h3}$, $\sum_{i=1}^{K}\sum_{j=1}^{K}\sum_{h=1}^{K}f_{ijh}p_{ijh}=-B$, $\sum_{i=1}^{K}\sum_{j=1}^{K}\sum_{h=1}^{K}f_{ijh}^{2}p_{ijh}=B(3X-1)/(2X-1)$ -which in general will be $B(RX-1)/[(R-1)X-1]$- and

$$V(\hat{\Delta})=\frac{1-\Delta}{n}\left\{\Delta+\frac{X}{(R-1)X-1}\right\}. \qquad \textbf{(C5)}$$

If $\hat{f}=\hat{\alpha}_t$ then $f=p_{ttt}-\lambda_t$ and $\Delta'_{t(ijh)}=f_{ijh}=\partial\Delta/\partial p_{ijh}=\delta_{ijh}-\lambda'_{t(ijh)}+\lambda'_{t(B)}\Delta'_{ijh}$. Substituting the derivatives, which are known from the previous paragraph, for $R=3$,

$$\Delta'_{t(ijh)}=\delta_{ijht}+\left(1-\delta_{ijh}\right)X_t\left[\left\{\frac{\delta_{it}}{\pi_{t1}}+\frac{\delta_{jt}}{\pi_{t2}}+\frac{\delta_{ht}}{\pi_{t3}}\right\}+\frac{(R-1)}{(R-1)X-1}\left\{1-\left(\frac{X_i}{\pi_{i1}}+\frac{X_j}{\pi_{j2}}+\frac{X_h}{\pi_{h3}}\right)\right\}\right]. \quad \textbf{(C6)}$$

By performing the necessary operations and generalizing this to any value of $R$, we obtain $\sum_{i=1}^{K}\sum_{j=1}^{K}\sum_{h=1}^{K}f_{ijh}p_{ijh}=\alpha_t$, $\sum_{i=1}^{K}\sum_{j=1}^{K}\sum_{h=1}^{K}f_{ijh}^2p_{ijh}=\alpha_t+BX_t[(R-1)X_t/\{(R-1)X-1\}-1]$. Through expression (C1), we obtain

$$V\left(\hat{\alpha}_t\right)=\frac{1}{n}\left[\alpha_t\left(1-\alpha_t\right)+\left(1-\Delta\right)X_t\left\{\frac{(R-1)X_t}{(R-1)X-1}-1\right\}\right]. \quad \textbf{(C7)}$$

Finally, if $\hat{f}=\hat{\mathcal{S}}_t$ then, by expression (B5), $f=\mathcal{S}_t=3\alpha_t/N_t$ and $f_{ijh}=\partial\mathcal{S}_t/\partial p_{ijh}=\{3/N_t\}\times[\Delta'_{t(ijh)}-\alpha_t(\delta_{ti}+\delta_{tj}+\delta_{th})/N_t]$ with $N_t=p_{t\bullet\bullet}+p_{\bullet t\bullet}+p_{\bullet\bullet t}=3p_t+D_t$. We obtain $\sum_{i=1}^{K}\sum_{j=1}^{K}\sum_{h=1}^{K}f_{ijh}p_{ijh}=0$ and $\sum_{i=1}^{K}\sum_{j=1}^{K}\sum_{h=1}^{K}f_{ijh}^2p_{ijh}=n\mathrm{V}(\hat{\mathcal{S}}_t)$ by expression (C1), with

$$V\left(\hat{\mathcal{S}}_t\right)=\frac{1}{n}\left(\frac{3}{N_t}\right)^2\left[\left\{\frac{2X_t}{2X-1}-1\right\}BX_t+\alpha_t-\frac{(2\times3-1)\alpha_t^2}{N_t}+2\left(\frac{\alpha_t}{N_t}\right)^2\left(p_{tt\bullet}+p_{t\bullet t}+p_{\bullet tt}\right)\right]. \quad \textbf{(C8)}$$

Generalizing to any value of $R$, $N_t=Rp_t+D_t=R\alpha_t+(1-\Delta)\sum_{r=1}^{R}\pi_{ir}$, and

$$V\left(\hat{\mathcal{S}}_t\right)=\frac{R^2}{nN_t^2}\left[\begin{array}{l}BX_t\left\{\dfrac{(R-1)X_t}{(R-1)X-1}-1\right\}+\alpha_t\left(1-\mathcal{S}_t\right)\left\{1-\dfrac{R-1}{R}\mathcal{S}_t\right\}+\\ +B\left(\dfrac{\mathcal{S}_t}{R}\right)^2\left\{\left(\sum_{r=1}^{R}\pi_{tr}\right)^2-\left(\sum_{r=1}^{R}\pi_{tr}^2\right)\right\}\end{array}\right]. \quad \textbf{(C9)}$$

The variances obtained in expressions (C5), (C7) and (C9) must be estimated based on the data of the problem; hence the expressions (13), (14) and (15), respectively.

**D. Variances in the special case of *R*=*K*=2**

As indicated in Section 4.2, the observed proportions verify the equalities

$\overline{p}_{i3} = \overline{p}_{3j} = \overline{p}_{33}$; thus, $\overline{d}_{11} = \overline{d}_{22}$, $\overline{d}_{12} = \overline{d}_{21}$ and $\overline{d}_{31} = \overline{d}_{32} = 2\overline{p}_{33}$. If we take the first expression (10), we obtain $\hat{\pi}_{11} = \hat{\pi}_{22}$, $\hat{\pi}_{12} = \hat{\pi}_{21}$ and $\hat{\pi}_{31} = \hat{\pi}_{32}$, so $\hat{X}_1 = \hat{X}_2$ and $\hat{X}_3 = \hat{\pi}_{31}^2 / \left(2\hat{\pi}_{31} - 1\right)$. Determining $\hat{V}\left(\hat{\mathcal{S}}_i\right)$ does not pose any problem, as the consistency $\hat{\mathcal{S}}_i$ in this case is defined in the same manner as it is defined in Section 4.1. Therefore, its variance is similar to the variance of expression (15) for $R$=2, that is, the variance of expression (18).

If $\hat{f} = \hat{\alpha}_t^*$, then $f=\alpha_t^* = \alpha_t/(1-p_{3\bullet})$ and $f_{ij}=(\partial f/\partial \alpha_t)\times(\partial \alpha_t/\partial p_{ij})+(\partial f/\partial p_{3\bullet})(\partial p_{3\bullet}/\partial p_{ij})=$ $[\partial \alpha_t/\partial p_{ij}+\alpha_t^*\delta_{i3}]/(1-p_{3\bullet})$, with $\partial \alpha_t/\partial p_{ij}=\Delta'_{t(ij)}$ given by expression (C6) for $R$=2 and omitting the subscript and terms in $h$. By performing the operations $\sum_{i=1}^{K}\sum_{j=1}^{K} f_{ij}p_{ij} = \alpha_t^*/(1-p_{3\bullet})$, $\sum_{i=1}^{K}\sum_{j=1}^{K} f_{ij}^2 p_{ij} = [(1-\Delta)X_t\{X_t/(X-1)-1\}+(1-p_{3\bullet})\alpha_t^*(1-\alpha_t^*)+\left(\alpha_t^*\right)^2]/(1-p_{3\bullet})^2$ and, through expression (C1), we obtain the following variance that yields the second expression (16)

$$V\left(\hat{\alpha}_t^*\right) = \frac{1}{n\left(1-p_{3\bullet}\right)^2}\left[\left(1-\Delta\right)X_t\left\{\frac{X_t}{X-1}-1\right\}+\left(1-p_{3\bullet}\right)\alpha_t^*\left(1-\alpha_t^*\right)\right].$$

If $\hat{f} = \hat{\Delta}^*$, then $f=\Delta^*=(\Delta-\alpha_3)/(1-p_{3\bullet})$ and $f_{ij}=(\partial f/\partial \Delta)\times(\partial \Delta/\partial p_{ij})+(\partial f/\partial \alpha_3)\times(\partial \alpha_3/\partial p_{ij})+$ $(\partial f/\partial p_{3\bullet})(\partial p_{3\bullet}/\partial p_{ij})=[\Delta'_{ij}-\Delta'_{3(ij)}+\Delta^*\delta_{i3}]/(1-p_{3\bullet})$, where $\Delta'_{ij}$ and $\Delta'_{3(ij)}$ are given by expressions (C4) and (C6), respectively, for $R$=2 and $t$=3. By performing the operations $\sum_{i=1}^{K}\sum_{j=1}^{K} f_{ij}p_{ij} =$ $-(1-\Delta^*)/(1-p_{3\bullet})$, $\sum_{i=1}^{K}\sum_{j=1}^{K} f_{ij}^2 p_{ij} = [\{B(1-X_3)(X-X_3)/(X-1)\}+(1-\Delta^*)(1-\Delta^* p_{3\bullet})]/(1-p_{3\bullet})^2$ and, through expression (C1), we obtain the following variance that yields expression (17)

$$\mathrm{V}\left(\hat{\Delta}^*\right) = \frac{1}{n\left(1-p_{3\bullet}\right)^2}\left[\left(1-\Delta\right)\left(1-X_3\right)\frac{X-X_3}{X-1}+\left(1-p_{3\bullet}\right)\Delta^*\left(1-\Delta^*\right)\right].$$

**Table 1**

**Diagnosis of $n$=100 subjects by $R$=2 raters in $K$=3 categories (Fleiss *et al.* 2003)**

**(a) Observed proportions ( $\bar{p}_{ij}$ )**

| | *Rater 2* | | | *Totals* |
|---|---|---|---|---|
| *Rater 1* | *Psychotic* | *Neurotic* | *Organic* | ( $\bar{p}_{i\bullet}$ ) |
| *Psychotic* | .75 | .01 | .04 | .80 |
| *Neurotic* | .05 | .04 | .01 | .10 |
| *Organic* | .00 | .00 | .10 | .10 |
| *Totals* ( $\bar{p}_{\bullet j}$ ) | .80 | .05 | .15 | 1 ( $\bar{p}_{\bullet\bullet}$ ) |

**(b) Data needed to make inferences with the *multi-rater delta* model. The data (raw) are obtained based on the data in Table 1(a)**

| *Categories* ($i$) | *Agreements* $\bar{p}_i$ | *Disagreements of rater r in category i* $\bar{d}_{ir}$ | | *Totals* $\bar{D}_i$ |
|---|---|---|---|---|
| | | *Rater=1* | *Rater=2* | |
| *Psychotic (1)* | .75 | .05 | .05 | .10 |
| *Neurotic (2)* | .04 | .06 | .01 | .07 |
| *Organic (3)* | .10 | .00 | .05 | .05 |
| *Totals* | $\bar{p}$ =.89 | $\bar{D}$ =.11 | $\bar{D}$ =.11 | 2 $\bar{D}$ =.22 |

**(c) Estimate of the parameters and the measures of the degree of agreement of the *multi-rater delta* model for the data in Table 1(b). In the case of measures of degree of agreement, it also indicates their SEs**

| *Categories* ($i$) | *Parameters* $\hat{\alpha}_i$ | *Parameters* $\hat{\pi}_{ir}$ | | *Consistencies* $\left(\hat{\mathcal{S}}_i \pm SE\right)$ |
|---|---|---|---|---|
| | | *r=1* | *r=2* | *(degree of agreement in class i)* |
| *Psychotic (1)* | .5500 | .8000 | .8000 | .6875±.1442* |
| *Neurotic (2)* | .0375 | .2000 | .0400 | .5000±.2058* |
| *Organic (3)* | .1000 | .0000 | .1600 | .8000±.1085* |
| *Overall degree of agreement* ( $\hat{\Delta}$ ±*SE*) | | | | .6875±.1099* |

* At least one estimation of $\hat{\pi}_{ir}$ is zero. Therefore, the values of *SE* have been obtained adding +.5 to all the observations $x_{ij}$.

**Table 2**

**Cognitive response cross-classification of *n*=164 subjects by *R*=3 raters in *K*=3 categories (Dillon and Mulani, 1984, p.449)**

**(a) Absolute frequencies $x_{i_1 i_2 i_3}$. Observed proportions are $\overline{p}_{i_1 i_2 i_3} = x_{i_1 i_2 i_3} / n$**

| *Rater 3* | | *1* | | | *2* | | | *3* | | |
|---|---|---|---|---|---|---|---|---|---|---|
| *Rater 2* | | *1* | *2* | *3* | *1* | *2* | *3* | *1* | *2* | *3* |
| | *1* | 56 | 1 | 0 | 5 | 3 | 0 | 0 | 0 | 1 |
| *Rater 1* | *2* | 12 | 2 | 1 | 14 | 20 | 4 | 0 | 4 | 2 |
| | *3* | 1 | 1 | 0 | 2 | 1 | 7 | 2 | 1 | 24 |

**(b) Data needed to apply the *multi-rater delta* model, which are obtained from Table 2(a)**

| *Categories* (*i*) | *Agreements* $n\,\overline{p}_i$ | *Disagreements of rater r in category i* $n\,\overline{d}_{ir}$ | | | *Total disagreements* $n\overline{D}_i$ |
|---|---|---|---|---|---|
| | | *Rater=1* | *Rater=2* | *Rater=3* | |
| *1* | 56 | 10 | 36 | 18 | 64 |
| *2* | 20 | 39 | 13 | 36 | 88 |
| *3* | 24 | 15 | 15 | 10 | 40 |
| *Totals* | $n\,\overline{p}$ =100 | $n\,\overline{D}$ =64 | $n\,\overline{D}$ =64 | $n\,\overline{D}$ =64 | $nR\,\overline{D}$ =192 |

**(c) Estimate of the parameters and the measures of the degree of agreement of the *multi-rater delta* model for the data in Table 2(b). In the case of measures of degree of agreement, it also indicates their SEs**

| *Categories* *(i)* | *Parameters* $\hat{\alpha}_i$ | *Parameters* $\hat{\pi}_{ir}$ | | | *Consistencies* $\left(\hat{\mathcal{S}}_i \pm SE\right)$ *(degree of agreement in class i)* |
|---|---|---|---|---|---|
| | | *r=1* | *r=2* | *r=3* | |
| *1* | .3320 | .1564 | .5084 | .2647 | .7040±.0460 |
| *2* | .0741 | .6343 | .2823 | .5937 | .2462±.1011 |
| *3* | .1435 | .2093 | .2093 | .1416 | .6306±.0668 |
| *Overall degree of agreement* ( $\hat{\Delta}$±*SE*) | | | | | .5496±.0462 |

**(d) Data needed to estimate Hubert's *kappa* (*R*-wise), which are obtained from Table 2(b)**

| *Categories* ($i$) | *Agreements* $n\,\overline{p}_i$ | *Number of responses i from rater r* $n\,\overline{t}_{ir} = n\left(\overline{p}_i + \overline{d}_{ir}\right)$ | | |
|---|---|---|---|---|
| | | *Rater=1* | *Rater=2* | *Rater=3* |
| *1* | 56 | 66 | 92 | 74 |
| *2* | 20 | 59 | 33 | 56 |
| *3* | 24 | 39 | 39 | 34 |
| *Totals* | $n\,\overline{p}$ =100 | $n$=164 | $n$=164 | $n$=164 |

**(e) Data needed to estimate Fleiss' *kappa*, which are obtained from Table 2(a)**

| *Categories* ($i$) | *Number of subjects* $s_{i\omega}$ *in which* $R_{si} = \omega$ *raters respond i (the value* $\omega$*=0 has no interest)* | | | *Total* $R_{\bullet i}$ *of responses i (all raters)* |
|---|---|---|---|---|
| | $\omega$=1 | $\omega$=2 | $\omega$=3=$R$ | $(\Sigma_\omega \omega s_{i\omega})$ |
| *1* | 26 | 19 | 56 | 232= $R_{\bullet 1}$ |
| *2* | 32 | 28 | 20 | 148= $R_{\bullet 2}$ |
| *3* | 14 | 13 | 24 | 112= $R_{\bullet 3}$ |
| *Totals* ($s_{\bullet\omega}$) | 72 | 60 | 100 | 492= $\sum_{i=1}^{3} R_{\bullet i} = \sum_{\omega=1}^{3} \omega s_{\bullet\omega}$ |

**(f) Summary of the different measures of degree of agreement for the data in Table 2(a)**

| | |
|---|---|
| *Raw* | .6098 |
| *Multi-rater delta* | .5496 |
| *Hubert's kappa* (*R-wise*) | .5471 |
| *Hubert's kappa* (*pairwise*) | .5809 |
| *Fleiss' kappa* | .5777 |

**Table 3**

**Modification of the data in Table 2 to obtain unbalanced marginal distributions ($n$=164)**

**(a) Absolute frequencies $x_{i_1 i_2 i_3}$. Observed proportions are $\bar{p}_{i_1 i_2 i_3} = x_{i_1 i_2 i_3} / n$**

| *Rater 3* | | *1* | | | *2* | | | *3* | | |
|---|---|---|---|---|---|---|---|---|---|---|
| *Rater 2* | | *1* | *2* | *3* | *1* | *2* | *3* | *1* | *2* | *3* |
| | *1* | 108 | 1 | 0 | 2 | 3 | 0 | 0 | 0 | 1 |
| *Rater 1* | *2* | 2 | 2 | 1 | 4 | 10 | 4 | 0 | 4 | 0 |
| | *3* | 2 | 1 | 0 | 7 | 1 | 2 | 4 | 1 | 4 |

**(b) Data needed to apply the *multi-rater delta* model, which are obtained from Table 3(a)**

| *Categories* ($i$) | *Agreements* $n\,\bar{p}_i$ | *Disagreements of rater r in category i* $n\,\bar{d}_{ir}$ | | | *Total disagreements* $n\bar{D}_i$ |
|---|---|---|---|---|---|
| | | *Rater=1* | *Rater=2* | *Rater=3* | |
| *1* | 108 | 7 | 21 | 9 | 37 |
| *2* | 10 | 17 | 13 | 23 | 53 |
| *3* | 4 | 18 | 8 | 10 | 36 |
| *Totals* | $n\,\bar{p}$ =122 | $n\,\bar{D}$ =42 | $n\,\bar{D}$ =42 | $n\,\bar{D}$ =42 | $nR\,\bar{D}$ =126 |

**(c) Data needed to estimate Hubert's *kappa* (*R*-wise), which are obtained from Table 3(b)**

| *Categories* ($i$) | *Agreements* $n\,\bar{p}_i$ | *Number of responses i from rater r* $n\,\bar{t}_{ir} = n\left(\bar{p}_i + \bar{d}_{ir}\right)$ | | |
|---|---|---|---|---|
| | | *Rater=1* | *Rater=2* | *Rater=3* |
| *1* | 108 | 115 | 129 | 117 |
| *2* | 10 | 27 | 23 | 33 |
| *3* | 4 | 22 | 12 | 14 |
| *Totals* | $n\,\bar{p}$ =122 | $n$=164 | $n$=164 | $n$=164 |

**(d) Data needed to estimate Fleiss' *kappa*, which are obtained from Table 3(a)**

| *Categories* ($i$) | *Number of subjects $s_{i\omega}$ in which $R_{si} = \omega$ raters respond i (the value $\omega$=0 has no interest)* | | | *Total $R_{\bullet i}$ of responses i (all raters)* |
|---|---|---|---|---|
| | $\omega$=1 | $\omega$=2 | $\omega$=3=$R$ | ($\Sigma_\omega \omega s_{i\omega}$) |
| *1* | 23 | 7 | 108 | 361=$R_{\bullet 1}$ |
| *2* | 17 | 18 | 10 | 83=$R_{\bullet 2}$ |
| *3* | 20 | 8 | 4 | 48=$R_{\bullet 3}$ |
| *Totals* ($s_{\bullet\omega}$) | 60 | 33 | 122 | 492=$\Sigma_i R_{\bullet i} = \Sigma_\omega \omega s_{\bullet\omega}$ |

**(e) Summary of the different measures of degree of agreement for the data in Table 3(a). Unbalanced marginals do not affect the *multi-rater delta* coefficient but do affect the *kappa* coefficient**

| | |
|---|---|
| *Raw* | .7439 |
| *Multi-rater delta* | .7075 |
| *Hubert's kappa* (*R-wise*) | .5739 |
| *Hubert's kappa* (*pairwise*) | .5553 |
| *Fleiss' kappa* | .5538 |

## Supplementary Material

**Paper:**
**Multi-rater delta: extending the delta nominal measure of agreement between two raters to many raters**
*Martín Andrés, A. and Álvarez Hernández, M.*

## SM1. Practical way for resolving equation (11) in the paper

To obtain the solutions to the general expression (11) we must perform the following steps for the categories with $\lambda_i \neq 0$. First, we must determine some of the quantities that will be needed and identify the behaviors of certain functions. In particular, the function $h_i(\lambda_i)=\prod_{r=1}^{R}\left(\lambda_i+\bar{d}_{ir}\right)/\lambda_i$ tends to infinity when $\lambda_i$ tends to zero or infinity. The minimum $\lambda_{i0}$ is attained at $\min_r \bar{d}_{ir}/\left(R-1\right)\le\lambda_{i0}\le\max_r \bar{d}_{ir}/\left(R-1\right)$, which is given by the solution of expression $\sum_{i=1}^{K}\lambda_i/(\lambda_i+\bar{d}_{ir})=1$. Let $B_i=[h_i(\lambda_{i0})]^{1/(R-1)}$. Given a certain value of $B$, the $\lambda_i$ values must verify that $h_i(\lambda_i)=B^{R-1}$, which implies that $B\ge B_t=\max_i B_i$ (where $t$ is the category in which the maximum occurs) for a solution to exist, in which case the $\lambda_i$ solution is double: $\lambda_i^- \le \lambda_{i0} \le \lambda_i^+$. Only the following quantities are needed: $\lambda_i^-$ ($\forall i \neq t$), $\lambda_t^-$ and $\lambda_t^+$, all of which are a function of $B$. The solutions have to be found within the limits $\prod_{r=1}^{R}\bar{d}_{ir}/B^{R-1}\le\lambda_i^-\le\lambda_{i0}$ ($\forall i$) and $\lambda_{t0}\le\lambda_t^+\le B-\min_r \bar{d}_{tr}$. When $B=B_t$, $\lambda_t^-=\lambda_{t0}=\lambda_t^+$. Note that the derivative with respect to $B$ of expression $h_i(\lambda_i)=B^{R-1}$ indicates that $\mathrm{d}\lambda_i^{\pm}/\mathrm{d}B=\left\{\prod_{r=1}^{R}\left(\lambda_i+\bar{d}_{ir}\right)/\lambda_i\right\}^{1/(R-1)}$ and $\lambda_i^-$ ($\lambda_i^+$) is decreasing (increasing) in $B$. When $B$ tends to infinity, $\lambda_i^-$ ($\lambda_i^+$) tends to zero (infinity) and $\lambda_i^+-B=\lambda_i^+-\left\{\prod_r\left(\lambda_i+\bar{d}_{ir}\right)/\lambda_i\right\}^{1/(R-1)}$ tends to $-\bar{D}_i/\left(R-1\right)$. Once these prior data are known the solution, which must be unique if it exists, is obtained as follows:

1) Let it be the function $g^-(B)=\sum_{i=1}^{K}\lambda_i^-+\bar{D}-B$, which is decreasing in $B$ –since $\mathrm{d}g^-/\mathrm{d}B=\sum_{i=1}^{K}\left(\mathrm{d}\lambda_i^-/\mathrm{d}B\right)-1<0$– and tends to minus infinity when B tends to infinity, since it tends to $0-\infty+\bar{D}$. Therefore if $g^{-}(B_t)=\sum_{i=1}^{K}\lambda_i^{-}+\bar{D}-B_t>0$, the solution $B$ (and $\lambda_i^{-}$ provided by it) must be found in $g^{-}(B)=\sum_{i=1}^{K}\lambda_i^{-}+\bar{D}-B=0$ for $B_t\leq B\leq\bar{D}+\sum_{i=1}^{K}\lambda_{i0}$.

2) When $g^{-}(B_t)=\sum_{i=1}^{K}\lambda_i^{-}+\bar{D}-B_t<0$, in which case the expression $g^{-}(B)=\sum_{i=1}^{K}\lambda_i^{-}+\bar{D}-B=0$ has no solution in $B$, we must consider the function $g_t^{+}(B)=\sum_{i=1,i\neq t}^{K}\lambda_i^{-}+\lambda_t^{+}+\bar{D}-B$. As $g_t^{+}(B_t)=g^{-}(B_t)<0$ and when $B$ tends to infinity $g_t^{+}(B)$ tends to $0+\bar{D}+[\lambda_t^{+}-B]=\bar{D}-\bar{D}_t/(R-1)\geq 0$, $B$ is the solution of expression $g_t^{+}(B)=0$, which has to be found in $B_t\leq B\leq\sum_{r=1}^{R}\sum_{r'=1}^{r}\left(\bar{D}-\bar{d}_{tr}\right)\left(\bar{D}-\bar{d}_{tr'}\right)/\left\{(R-1)\bar{D}-\bar{D}_t\right\}$.

A particular case is when $(R-1)\bar{D}=\bar{D}_i$, in which case $B=\infty$, $\lambda_i^{-}=0$, $\lambda_t^{+}=\infty$, $\hat{\alpha}_i=\bar{p}_i$ $(\forall i\neq t)$, $\hat{\alpha}_t=-\infty$, $\hat{\Delta}=-\infty$, $\hat{\pi}_{ir}=0$ $(\forall i\neq t)$ and $\hat{\pi}_{tr}=1$; in this circumstance, the problem should be solved by increasing the original frequencies by 0.5 (refer to the subsequent section SM2).

The reason for the last statement is that the total agreement obtained, $\hat{\Delta}=-\infty$, is misleading because it is the result of a bad agreement in category $t$ ($\hat{\alpha}_t=-\infty$), which dominates the agreements in the remainder of the categories ($\hat{\alpha}_i=\bar{p}_i$).The model fits poorly with the data due to the strange behavior of category $t$, which can be corrected by adding 0.5 to the data to avoid zero frequencies.

3) If $g^{-}(B_t)=0$ then $B=B_t$ is solution of $g^{-}(B)=0$ and of $g_t^{+}(B)=0$. But, as indicated in 1) and 2), the solution is unique if $(R-1)\bar{D}>\bar{D}_t$. When $(R-1)\bar{D}=\bar{D}_t>0$, any value $B\geq B_t$ is solution and the problem has to be solved by increasing the original frequencies by 0.5. In this case, there are infinite solutions because there are more unknown parameters than free cells to take values. When $(R-1)\bar{D}=\bar{D}_t=0$, it is the special case of $\bar{D}=0$: $\lambda_i=0$ $(\forall i)$, $\hat{\alpha}_i=\bar{p}_i$, $\hat{\Delta}=1$ and $\hat{\pi}_{ir}$ values are indeterminate.

**SM2. Equivalence of the estimators of the *multi-rater delta* and classical *delta* model parameters when *R*=2**

When $R$=2, the first expression of (11) is $B$=$(\lambda_i+\bar{d}_{i1})(\lambda_i+\bar{d}_{i2})/\lambda_i$. Therefore, $2\lambda_i=B-(\bar{d}_{i1}+\bar{d}_{i2})+s(i)R_i$, with $s(i)=\pm1$ and $R_i^2=B^2-2B\left(\bar{d}_{i1}+\bar{d}_{i2}\right)+\left(\bar{d}_{i1}-\bar{d}_{i2}\right)^2$. By the second expression of (11), the condition for determining the value of $B$ is $h(B)$= $(K-2)B+\sum_{i=1}^{K}s\left(i\right)R_i$ =0. Given the value of $B$, the parameter estimates of the *multi-rater delta* model are obtained by expressions (10); for example,

$$\hat{\pi}_{i1}=\frac{B+\left(\bar{d}_{i1}-\bar{d}_{i2}\right)+s\left(i\right)R_i}{2B},\hat{\pi}_{i2}=\frac{B+\left(\bar{d}_{i2}-\bar{d}_{i1}\right)+s\left(i\right)R_i}{2B}\text{ and }\hat{\alpha}_i=\frac{\bar{p}_{i\bullet}+\bar{p}_{\bullet i}-B+s\left(i\right)R_i}{2}.$$

It is easy to see that the parameters $\lambda_{i0}$ defined in the previous section are $\lambda_{i0}=\sqrt{\bar{d}_{i1}\bar{d}_{i2}}$; therefore $B_i=\left(\sqrt{\bar{d}_{i1}}+\sqrt{\bar{d}_{i2}}\right)^2$. Note that now it is not necessary to control when $\lambda_i$=0 occurs, since the expressions themselves detect it; for example, if $\bar{d}_{i2}$=0, then $R_i=\left|B-\bar{d}_{i1}\right|=(B-\bar{d}_{i1})$ since $B\geq B_i=\bar{d}_{i1}$ and $\hat{\pi}_{i2}$=0 in $s(i)=-1$. The consequence is that the *delta* classic and *multi-rater delta* models provide the same estimates, as deduced from expressions (11).

Regarding the determination of the value of B, Martín Andrés & Femia Marzo (2004 and 2005) were mistaken in their assertion that the problem has no solution or infinite solutions, which are related to the presence of $\bar{d}_{ir}=0$ or the solution $B$=∞. As shown in the previous section, a solution is also obtained in these cases, but it implies that $\lambda_i$=0 and $\hat{\pi}_{ir}$=0. As indicated in 2) of the previous section, the case of $B$=∞ can occur only when $\bar{D}=\bar{D}_t$ (that is, when all disagreements are concentrated in class $t$). The Martín Andrés & Femia Marzo instructions should be corrected for their *delta* model. The correct instructions are given below with a notation similar to the notation of Martín Andrés & Femia Marzo, whose calculations are based on the observed frequencies $x_{ij}$ instead of the observed proportions $\overline{p}_{ij}$:

If $K>2$, determine the only solution $\tilde{B}=n(1-\hat{\Delta})\geq\tilde{B}_t=\max_i\left\{\sqrt{x_{\bullet i}-x_{ii}}+\sqrt{x_{i\bullet}-x_{ii}}\right\}^2$ of the equation

$$h(\tilde{B})=(K-2)\tilde{B}+\sum_{i=1}^{K}s(i)\sqrt{\left\{\tilde{B}+\left(x_{\bullet i}-x_{i\bullet}\right)\right\}^2-4\tilde{B}\left(x_{\bullet i}-x_{ii}\right)}=0$$

where $s(i)=-1$ always, except when in such values it occurs that $h(\tilde{B}_t)<0$, in which case $s(t)=+1$ in the category $t$ that provides the $\tilde{B}_t$ value. Once $\tilde{B}$ is calculated,

$$\hat{\pi}_i=\frac{\tilde{B}+\left(x_{\bullet i}-x_{i\bullet}\right)+s(i)\sqrt{\left\{\tilde{B}+\left(x_{\bullet i}-x_{i\bullet}\right)\right\}^2-4\tilde{B}\left(x_{\bullet i}-x_{ii}\right)}}{2\tilde{B}},\quad x_{i\bullet}\hat{\Delta}_i=\frac{x_{ii}-x_{i\bullet}\hat{\pi}_i}{1-\hat{\pi}_i}\ \text{and}\ \hat{\Delta}=1-\frac{\tilde{B}}{n}.$$

In the previous calculations, three special cases can occur:

**a)** If $n-\sum_{i=1}^{K}x_{ii}=0$, that is, if no disagreements occur, then the solution is $\tilde{B}=0$, and therefore $\hat{\pi}_i$=--- (indeterminate), $x_{ii}\hat{\Delta}_i=x_{ii}$ and $\hat{\Delta}=1$.

**b)** In another case, if $h(\tilde{B}_t)<0$ and $n-\sum_{i=1}^{K}x_{ii}=x_{\bullet t}+x_{t\bullet}-2x_{tt}$, that is, if all disagreements are concentrated in the category $t$ that provides the $\tilde{B}_t$ value, then the solution is $\tilde{B}_t=\infty$, and therefore $\hat{\pi}_i=0$ ($\forall i\neq t$), $\hat{\pi}_t=1$, $x_{i\bullet}\hat{\Delta}_i=x_{ii}$ ($\forall i\neq t$), and $\hat{\Delta}_t=-\infty$. In this case, adding 0.5 to all data $x_{ij}$ to find the new solution is more convenient.

**c)** In another case, if $h(\tilde{B}_t)=0$ and $n-\sum_{i=1}^{K}x_{ii}=x_{\bullet t}+x_{t\bullet}-2x_{tt}>0$, that is, if there are disagreements and all of them are concentrated in the category $t$ that provides the $\tilde{B}_t$ value, then there are infinite $\tilde{B}$ solutions and the problem must be solved by adding 0.5 to all of data $x_{ij}$.

In the previous instructions, the explicit expression of $\hat{\Delta}_i$ has not been employed to anticipate any special circumstance. For example, in case a), $x_{ii}\hat{\Delta}_i$=$x_{ii}$ has been written; therefore, $\hat{\Delta}_i$=1 when $x_{ii}$≠0 but is indeterminate when $x_{ii}$=0.

As an example of case b), consider the data in Table 5 of Martín Andrés & Femia Marzo (2004). The absolute frequencies of their 3×3 table written by rows are 75, 1, 0 (1st row); 5, 4, 0 (2nd row); and 0, 1, 10 (3rd row). For these authors, the classic *delta* model does not yield a solution, which explains why they increased the data by 0.5 and obtained the value $\hat{\Delta}$=0.811. As indicated by the new instructions, the problem has a solution, obtaining the values $B$=∞, $\lambda_1$=0, $\lambda_2$=∞, and $\lambda_3$=0, the total degree of agreement of $\hat{\Delta}$=–∞, and consistencies $\hat{\mathcal{S}}_1$=0.9615, $\hat{\mathcal{S}}_2$=–∞, and $\hat{\mathcal{S}}_3$=0.9524. The total degree of agreement is very poor (–∞), which is caused by the value $\hat{\mathcal{S}}_2$=–∞, as the degree of agreement (consistency) in categories 1 and 3 is quite good. A good option to evaluate the total agreement is to add 0.5 to the data, which yields the value 0.811 of Martín Andrés & Femia Marzo. Even more remarkable is the case that in the previous table only the two data $x_{12}$=$x_{32}$=1 are changed by $x_{12}$=$x_{32}$=0; thus, $n$=94. According to the instructions of Martín Andrés & Femia Marzo (2005), the problem has infinite solutions, and 0.5 must be added to all data; a value of $\hat{\Delta}$=0.8780 is obtained. From the perspective of the *multi-rater delta* model, the problem does have a solution that is very simple to obtain. Since $\bar{d}_{11} = \bar{d}_{22} = \bar{d}_{31} = 0$, $\lambda_1$=$\lambda_2$=$\lambda_3$=0, and, by the last expression of (10), $B$=$\bar{D}$=5/94; thus, $\hat{\Delta}$=89/94=0.9468.

Once demonstrated that the estimators by one or the other model are the same, it is convenient to demonstrate that their variances are also. The following steps demonstrate the case of $\hat{V}\left(\hat{\Delta}\right)$; the other cases are checked in a similar manner. According to Martín Andrés & Femia Marzo (2005), $\hat{V}\left(\hat{\Delta}\right)$=$(n–E^{-1}–n\hat{\Delta}^2)/n^2$, with $E$=$\sum_{i=1}^{K} E_i$, $E_i$= $\hat{\pi}_i \Big/ \left\{ n\left(1-\hat{\Delta}\right) - x_{i\bullet} v_i \right\}$ and $v_i = \left(1-\hat{\Delta}_i\right) \Big/ \left(1-\hat{\pi}_i\right)$. Substituting values gives $E$=

$\sum_{i=1}^{K}\left[\hat{\pi}_i\left(1-\hat{\pi}_i\right)\Big/\left\{n\left(1-\hat{\Delta}\right)\left(1-\hat{\pi}_i\right)-x_{i\bullet}\left(1-\hat{\Delta}_i\right)\right\}\right]$. Considering that the expressions (A5), which are expressed for the estimates, indicate that $\hat{\pi}_i=\hat{\pi}_{i2}$ and $x_{i\bullet}\left(1-\hat{\Delta}_i\right)=n\left(1-\hat{\Delta}\right)\hat{\pi}_{i1}$, then $E=[n(1-\hat{\Delta})]^{-1}\sum_{i=1}^{K}\left\{\hat{\pi}_{i2}\left(1-\hat{\pi}_{i2}\right)/\left(1-\hat{\pi}_{i1}-\hat{\pi}_{i2}\right)\right\}=[n(1-\hat{\Delta})]^{-1}[1+\sum_{i=1}^{K}\left\{\hat{\pi}_{i1}\hat{\pi}_{i2}/\left(1-\hat{\pi}_{i1}-\hat{\pi}_{i2}\right)\right\}]=[n(1-\hat{\Delta})]^{-1}[1-\sum_{i=1}^{K}\hat{X}_i]=(1-\hat{X})/n(1-\hat{\Delta})$. Substituting gives $\hat{V}\left(\hat{\Delta}\right)=[(1-\hat{\Delta})/n][\hat{\Delta}+\hat{X}/(\hat{X}-1)]$, that is, the expression (13) for $R$=2.

Let us now suppose that $K$=2. As demonstrated at the end of Appendix D, these three variances are the same variances estimated by Martín Andrés & Femia Marzo (2005). Because these authors also consider the cases in which rater 1 is a standard and the marginal of the standard rater is fixed, Table SM1 shows the different possible parameters and their SEs based on the parameters of the current *multi-rater delta* model. This table is the same as Table 8 of Martín Andrés & Femia Marzo (2005), and that paper should be referred to in order to understand some of the concepts in Table SM1.

## Table SM1

**Measures of degree of agreement, using the notation of the *multi-rater delta* model, for *K* categories and two raters**

<table>
<tr><th><i>Name</i></th><th><i>K</i></th><th><i>Valid</i></th><th><i>Sampling</i></th><th><i>Estimation</i></th><th><i>Estimated variance</i></th></tr>
<tr><td rowspan="2"><i>Conformity</i> (in <i>i</i>)</td><td rowspan="4">≥2</td><td rowspan="3">If rater 1 is a standard rater</td><td>I</td><td rowspan="2">$\hat{\mathcal{F}}_i = \dfrac{\hat{\alpha}_i}{\overline{p}_{i\bullet}}$</td><td rowspan="2">$\dfrac{\hat{H}_i + \overline{p}_{i\bullet}\hat{\mathcal{F}}_i\left(1-\hat{\mathcal{F}}_i\right)}{n\overline{p}_{i\bullet}^2}$</td></tr>
<tr><td>II</td></tr>
<tr><td><i>Predictivity</i> (in <i>i</i>)</td><td>I</td><td>$\hat{\mathcal{P}}_i = \dfrac{\hat{\alpha}_i}{\overline{p}_{\bullet i}}$</td><td>$\left[\hat{H}_i + \overline{p}_{\bullet i}\hat{\mathcal{P}}_i\left(1-\hat{\mathcal{P}}_i\right)\right]\Big/ n\overline{p}_{\bullet i}^2$</td></tr>
<tr><td><i>Consistency</i> (in <i>i</i>)</td><td>If no standard rater exists</td><td>I</td><td>$\hat{\mathcal{S}}_i = \dfrac{2\hat{\alpha}_i}{\overline{p}_{i\bullet} + \overline{p}_{\bullet i}}$</td><td>$\dfrac{4}{n\overline{N}_i^2}\left[\hat{H}_i + \hat{\alpha}_i\left\{1-\dfrac{3\hat{\alpha}_i}{\overline{N}_i}+\dfrac{2\hat{\alpha}_i\overline{p}_{ii}}{\overline{N}_i^2}\right\}\right]$</td></tr>
<tr><td rowspan="4"><i>Agreement</i> (in <i>i</i>)</td><td rowspan="2">>2</td><td rowspan="4">Always</td><td>I</td><td rowspan="2">$\hat{\alpha}_i$</td><td>$\left[\hat{H}_i + \hat{\alpha}_i\left(1-\hat{\alpha}_i\right)\right]\Big/ n$</td></tr>
<tr><td>II</td><td>$\left[\hat{H}_i + \hat{\alpha}_i\left(1-\hat{\alpha}_i/\overline{p}_{i\bullet}\right)\right]\Big/ n$</td></tr>
<tr><td rowspan="2">=2</td><td>I</td><td rowspan="2">$\hat{\alpha}_i^* = \dfrac{\hat{\alpha}_i}{1-p_{3\bullet}}$</td><td>$\dfrac{\hat{H}_i + \left(1-p_{3\bullet}\right)\hat{\alpha}_i^*\left(1-\hat{\alpha}_i^*\right)}{n\left(1-p_{3\bullet}\right)^2}$</td></tr>
<tr><td>II</td><td>$\dfrac{\hat{H}_i + \hat{\alpha}_i\left\{1-\hat{\alpha}_i/\overline{p}_{i\bullet}\right\}}{n\left(1-p_{3\bullet}\right)^2}$</td></tr>
<tr><td rowspan="4"><i>Overall degree of agreement</i></td><td rowspan="2">>2</td><td rowspan="4">Always</td><td>I</td><td rowspan="2">$\hat{\Delta}$</td><td>$\left[\hat{H} + \hat{\alpha}\left(1-\hat{\alpha}\right)\right]\Big/ n$</td></tr>
<tr><td>II</td><td>$\left[\hat{H} + \sum\hat{\alpha}_i\left(1-\hat{\alpha}_i/\overline{p}_{i\bullet}\right)\right]\Big/ n$</td></tr>
<tr><td rowspan="2">=2</td><td>I</td><td rowspan="2">$\hat{\Delta}^* = \dfrac{\hat{\alpha}_1+\hat{\alpha}_2}{1-p_{3\bullet}} = \hat{\alpha}_1^* + \hat{\alpha}_2^*$</td><td>$\dfrac{\hat{H}^* + \left(1-\overline{p}_{3\bullet}\right)\hat{\Delta}^*\left(1-\hat{\Delta}^*\right)}{n\left(1-p_{3\bullet}\right)^2}$</td></tr>
<tr><td>II</td><td>$\dfrac{\hat{H}^* + \sum_{i=1}^{2}\hat{\alpha}_i\left(1-\hat{\alpha}_i/\overline{p}_{i\bullet}\right)}{n\left(1-p_{3\bullet}\right)^2}$</td></tr>
</table>

NOTES: **(1)** To understand the sense of the measures of the degree of agreement *Conformity* and *Predictivity* please refer to the original paper by Martín Andrés and Femia Marzo (2005); **(2)** If *K*=2, measures of degree of agreement are only defined for *i*=1 or 2; **(3)** in Sampling I, all marginal totals have been randomly obtained; in Sampling II, the marginal of rater 1 has been previously fixed; and **(4)** $\overline{N}_i = \overline{p}_{i\bullet} + \overline{p}_{\bullet i}$, $\hat{H}_i = \left(1-\hat{\Delta}\right)\hat{X}_i\left\{-1+\hat{X}_i/\left(\hat{X}-1\right)\right\}$, $\hat{H} = \left(1-\hat{\Delta}\right)\hat{X}/\left(\hat{X}-1\right)$ and $\hat{H}^* = \left(1-\hat{\Delta}\right)\left(\hat{X}-\hat{X}_3\right)\left(1-\hat{X}_3\right)/\left(\hat{X}-1\right)$.